\IfFileExists{lmodern.sty}{}{\PassOptionsToPackage{expansion=false}{microtype}}
\documentclass[a4paper,UKenglish,cleveref,autoref,thm-restate]{lipics-v2021}
\nolinenumbers
\usepackage{booktabs}
\usepackage{array}
\usepackage{tikz}
\usetikzlibrary{arrows.meta,positioning,fit}
\usepackage{pgfplots}
\pgfplotsset{compat=1.17}
\usepgfplotslibrary{groupplots}
\definecolor{vuBlue}{HTML}{0072B2}
\definecolor{vuRed}{HTML}{D55E00}
\definecolor{vuGreen}{HTML}{009E73}
\pgfplotsset{rnd/.style={black!55, densely dashed, semithick, mark=none},
  occl/.style={vuBlue, thick, mark=*, mark size=1.4pt},
  gxi/.style={vuRed, thick, mark=square*, mark size=1.3pt, densely dotted},
  ig/.style={vuGreen, thick, mark=triangle*, mark size=1.7pt, dashdotted}}

\newcommand{\NumFunctions}{515}
\newcommand{\NumPairs}{481}
\newcommand{\NumCommon}{481}
\newcommand{\NumSelfCommon}{376}
\newcommand{\NumCandidateFunctions}{672}
\newcommand{\NumExcludedFixedNotClean}{39}

\newcommand{\NumExcludedCveMatch}{106}

\newcommand{\NumExcludedRecordCheck}{12}

\newcommand{\RcBeyondPocMain}{109}

\newcommand{\FixedExecutedMain}{494}

\newcommand{\NumPatchDerived}{152}
\newcommand{\NumObservedTrigger}{363}
\newcommand{\NumProjects}{152}
\newcommand{\NumDistinctFunctions}{728}
\newcommand{\NumUnpaired}{34}

\newcommand{\NumSlicedInstances}{739}

\newcommand{\YearMin}{2005}
\newcommand{\YearMax}{2022}

\newcommand{\GptSixOneBothVuln}{54.9}

\newcommand{\GptFiveSixBothVuln}{49.3}

\newcommand{\CodeLlamaADR}{0.0}

\newcommand{\CodeLlamaFOne}{0.665}

\newcommand{\OnADRQwenThreeSix}{23.5}

\newcommand{\OnADRGptOss}{21.6}

\newcommand{\FirstBudgetADRQwenThreeSix}{24.5}

\newcommand{\FirstBudgetADRGptOss}{23.9}

\newcommand{\SurfaceADR}{5.8}
\newcommand{\SurfaceMCC}{0.046}
\newcommand{\SurfaceAUC}{0.530}

\newcommand{\SurfaceGapSixOneCI}{28.9 [23.9, 34.2]}

\newcommand{\SurfaceGapFiveSixCI}{20.2 [15.4, 25.3]}

\newcommand{\SurfaceGapClaudeCI}{61.3 [56.4, 66.0]}

\newcommand{\ChanceGapClaude}{43.4 [39.2, 47.4]}

\newcommand{\SurfaceGapClaudeFourEightCI}{33.9 [29.2, 39.1]}

\newcommand{\ChanceGapQwenThreeSix}{0.0 [$-$4.1, 4.0]}

\newcommand{\ChanceGapGptOss}{$-$3.4 [$-$7.3, 0.8]}

\newcommand{\ChanceGapFoundationSec}{1.3 [$-$1.6, 4.2]}

\newcommand{\ChanceGapVulnLLM}{$-$0.8 [$-$4.7, 3.0]}

\newcommand{\FixObsFunctions}{341}
\newcommand{\FixObsTrigChanged}{40.2}

\newcommand{\FixObsAllAfter}{29}

\newcommand{\SixOneKeywordGapObs}{45.7 [40.4, 51.4]}

\newcommand{\FiveSixKeywordGapObs}{38.2 [32.7, 43.7]}

\newcommand{\ClaudeKeywordGapObs}{55.7 [50.6, 61.0]}

\newcommand{\ClaudeFourEightKeywordGapObs}{33.3 [27.1, 39.7]}

\newcommand{\CitesObsFunctions}{342}
\newcommand{\JaccDminusJoernSixOne}{0.043 [0.026, 0.059]}
\newcommand{\JaccDminusDgSixOne}{$-$0.004 [$-$0.019, 0.011]}
\newcommand{\JaccDminusDwkSixOne}{0.071 [0.054, 0.087]}

\newcommand{\KeywordHasTrigger}{61.1}

\newcommand{\KeywordGapCIGptOss}{16.9 [12.2, 21.3]}

\newcommand{\KeywordGapCIQwenThreeSix}{21.0 [16.7, 25.3]}

\newcommand{\SixOneJaccD}{0.299}
\newcommand{\SixOneJaccW}{0.228}

\newcommand{\MaskDvsRandomSixOne}{$-$8.7 [$-$12.1, $-$5.4]}

\newcommand{\MaskDvsRandomFiveSix}{$-$9.8 [$-$13.4, $-$6.3]}

\newcommand{\MaskRandomVsNoneFiveSix}{7.5 [4.4, 10.7]}

\newcommand{\MaskDvsRandomClaude}{0.2 [$-$2.5, 2.9]}

\newcommand{\MaskYesNoneClaude}{94.0}
\newcommand{\MaskYesBlankDClaude}{93.1}

\newcommand{\MaskOrderedNoneClaude}{89.4}
\newcommand{\MaskOrderedBlankDClaude}{82.1}
\newcommand{\MaskOrderedRandomClaude}{84.0}
\newcommand{\MaskYesBlankDNamedClaude}{93.5}
\newcommand{\MaskYesBlankDUnnamedClaude}{90.1}
\newcommand{\MaskDvsRandomClaudeFourEight}{$-$4.6 [$-$8.4, $-$0.8]}

\newcommand{\MarkVvsPlainSixOne}{6.4 [2.9, 10.1]}

\newcommand{\MarkRandomvsPlainSixOne}{$-$2.7 [$-$5.7, 0.2]}
\newcommand{\MarkSvsRandomSixOne}{4.2 [0.8, 7.8]}
\newcommand{\EdgesVsShuffledSixOne}{4.6 [0.7, 8.7]}

\newcommand{\MarkVvsPlainFiveSix}{7.5 [3.6, 11.7]}

\newcommand{\MarkRandomvsPlainFiveSix}{$-$1.2 [$-$4.3, 2.1]}

\newcommand{\NestCorrect}{77.8 [71.1, 83.8]}
\newcommand{\NestWrong}{61.5 [56.5, 66.7]}

\newcommand{\LenShortBothVuln}{35.2}
\newcommand{\LenLongBothVuln}{81.1}
\newcommand{\LenShortADR}{47.9}
\newcommand{\LenLongADR}{18.9}
\newcommand{\ClaudeADR}{67.2}

\newcommand{\ClaudeBothVuln}{26.8}

\newcommand{\ClaudeJaccD}{0.315}

\newcommand{\LenShortBothVulnClaude}{15.8}
\newcommand{\LenLongBothVulnClaude}{43.2}

\newcommand{\TplAllThree}{2.3}
\newcommand{\TplAny}{17.7}

\newcommand{\HeadThirtyTwoAUCbefore}{0.518}
\newcommand{\HeadThirtyTwoAUCafter}{0.595}

\newcommand{\CohortSize}{150}

\newcommand{\RenameConfirmedFunctions}{96}

\newcommand{\VariantsConfirmed}{315}

\newcommand{\DecoyFaultFunctions}{20}
\newcommand{\DecoyCleanFunctions}{12}
\newcommand{\AltFixFunctions}{92}

\newcommand{\RankedTopOneMin}{3}
\newcommand{\RankedTopOneMax}{9}

\newcommand{\MaskADRNoneClaudeFourEight}{39.7}

\newcommand{\MaskADRRandomClaudeFourEight}{50.7}

\newcommand{\RankedCells}{76}
\newcommand{\RankedUp}{36}
\newcommand{\RankedDown}{15}

\newcommand{\RankedMissMin}{70.2}
\newcommand{\RankedMissMax}{88.1}

\newcommand{\KwTopOneMin}{10.3}
\newcommand{\KwTopOneMax}{11.0}
\newcommand{\AttrBelowKwTopOne}{17}
\newcommand{\AttrCombos}{19}

\newcommand{\VulnOcclJaccD}{0.207}
\newcommand{\VulnOcclKwJaccD}{0.194}

\newcommand{\CweUafCiteSixOne}{73.2}
\newcommand{\CweUafADRSixOne}{36.1}

\newcommand{\ExampleReadTrigger}{9}
\newcommand{\ExampleReadConfVuln}{98}
\newcommand{\ExampleReadConfFixed}{10}
\newcommand{\ExampleReadAnsVuln}{YES}
\newcommand{\ExampleReadAnsFixed}{NO}

\newcommand{\ExampleStaleTrigger}{29}
\newcommand{\ExampleStaleConfVuln}{100}

\newcommand{\ExampleStaleAnsVuln}{YES}

\newcommand{\SliceSizeDyn}{16.7}

\newcommand{\SliceRetJoern}{79.5}
\newcommand{\SliceContainJoern}{205}

\newcommand{\SliceRetDg}{97.4}
\newcommand{\SliceContainDg}{459}
\newcommand{\SliceSizeDg}{31.7}

\newcommand{\SliceRetDwk}{85.0}

\newcommand{\SliceSizeDwkPub}{57.8}

\newcommand{\SliceMissDwkPub}{19.0}
\newcommand{\SliceRetDwkPub}{79.4}

\newcommand{\SliceExtraMin}{32}
\newcommand{\SliceExtraMax}{43}

\newcommand{\SliceDgMinusJoern}{17.9 [15.7, 20.4]}

\newcommand{\CthreeASixOneMCC}{0.859}

\newcommand{\CthreeASixOneLine}{97.0}

\newcommand{\CthreeASixOneQuestions}{533}
\newcommand{\CthreeASixOneFunctions}{109}
\newcommand{\CthreeAAlwaysFault}{42.2}

\newcommand{\CthreeASixOneKeywordLine}{26.6}

\newcommand{\CthreeAFiveSixMCC}{0.826}

\newcommand{\CthreeAClaudeMCC}{0.891}

\newcommand{\CthreeAClaudeFourEightMCC}{0.734}

\newcommand{\CthreeAAccMin}{86.3}
\newcommand{\CthreeAAccMax}{94.6}

\newcommand{\GuardLocalAcc}{47.1}

\newcommand{\GuardLocalFaultShare}{82.7}
\newcommand{\GuardAddedAcc}{80.3}

\newcommand{\GuardAddedFzero}{67.0}
\newcommand{\GuardAddedDecoyFault}{15.0}

\newcommand{\CthreeAOverGuardMin}{6.0}
\newcommand{\CthreeAOverGuardMax}{14.3}
\newcommand{\CthreeAOverGuardClaudeFourEight}{6.0 [1.1, 10.7]}
\newcommand{\CthreeAOverGuardClaude}{14.3 [9.4, 19.2]}
\newcommand{\CthreeAOverGuardFiveSix}{10.9 [6.2, 15.7]}
\newcommand{\CthreeAOverGuardSixOne}{12.6 [7.5, 17.7]}

\newcommand{\PostCutCthreeAOverGuardMin}{$-$6.0}
\newcommand{\PostCutCthreeAOverGuardMax}{2.0}

\newcommand{\BenignFunctions}{60}
\newcommand{\BenignGenerated}{46}
\newcommand{\BenignMechanical}{14}

\newcommand{\BenignVzeroCleanMin}{98.3}
\newcommand{\BenignSwitchedMin}{68.3}
\newcommand{\BenignSwitchedMax}{88.3}
\newcommand{\CtwoSixOneFunctions}{109}

\newcommand{\CtwoSixOneRenameSame}{95.8}

\newcommand{\CtwoSixOneDevFixNo}{42.7}

\newcommand{\CtwoFiveSixRenameSame}{96.9}

\newcommand{\CtwoFiveSixDevFixNo}{32.5}

\newcommand{\CtwoClaudeRenameSame}{96.9}

\newcommand{\CtwoClaudeDevFixNo}{83.3}
\newcommand{\CtwoClaudeAltFixNo}{81.1}

\newcommand{\CtwoClaudeFourEightRenameSame}{87.5}

\newcommand{\CtwoClaudeFourEightDevFixNo}{58.1}
\newcommand{\CtwoClaudeFourEightAltFixNo}{63.5}

\newcommand{\RecSixOneProject}{95.9}
\newcommand{\RecSixOneCVE}{20.0}
\newcommand{\RecSixOneADRcve}{41.7 [32.0, 52.1]}
\newcommand{\RecSixOneADRprojectOnly}{32.9 [27.5, 38.3]}

\newcommand{\TcRule}{84.8}

\newcommand{\RecFiveSixCVE}{15.3}

\newcommand{\RecClaudeCVE}{68.5}
\newcommand{\RecClaudeADRcve}{72.1 [67.1, 77.0]}
\newcommand{\RecClaudeADRprojectOnly}{56.8 [48.6, 65.1]}

\newcommand{\RecClaudeFourEightCVE}{9.9}

\newcommand{\RecCthreeAClaudeDiff}{$-$1.4 [$-$6.4, 4.2]}

\newcommand{\DpsSixOneFlips}{35.1}

\newcommand{\DpsFiveSixFlips}{33.7}

\newcommand{\DpsClaudeFlips}{65.3}

\newcommand{\DpsClaudeFourEightFlips}{48.7}

\newcommand{\DpsClaudeFourEightNeutral}{87.3}

\newcommand{\DpsQwenSevenVuln}{12.2}
\newcommand{\DpsCommon}{205}
\newcommand{\IsoGptQueries}{47,996}

\newcommand{\PostCutFunctions}{97}
\newcommand{\PostCutConfirmed}{217}
\newcommand{\PostCutStudyCVEs}{195}

\newcommand{\PostCutClaudeVsSixOne}{7.2 [$-$5.2, 18.6]}

\newcommand{\PostCutAllClaudeVsSixOne}{2.1 [$-$6.7, 11.3]}

\newcommand{\PostCutDateFirst}{3 September 2026}
\newcommand{\PostCutDateLast}{9 November 2026}
\newcommand{\PostCutProjects}{100}
\newcommand{\PostCutCVELow}{CVE-2026-19445}
\newcommand{\PostCutCVEHigh}{CVE-2026-125405}
\newcommand{\PostCutFixFirst}{1 September 2026}
\newcommand{\PostCutFixLast}{5 November 2026}

\newcommand{\PostCutClaudeBlankVNoToYes}{37}
\newcommand{\PostCutClaudeBlankVNo}{48}

\newcommand{\YearRecentFunctions}{66}

\newcommand{\YearPostMinusRecentClaude}{$-$18.9 [$-$34.0, $-$3.2]}

\newcommand{\DidChangeSixOne}{$-$4.8 [$-$14.5, 5.3]}
\newcommand{\DidClaudeMinusSixOne}{$-$25.2 [$-$38.0, $-$12.0]}

\newcommand{\DidChangeFiveSix}{$-$1.2 [$-$10.6, 9.1]}
\newcommand{\DidClaudeMinusFiveSix}{$-$28.8 [$-$42.5, $-$15.6]}
\newcommand{\DidMainClaude}{67.2}
\newcommand{\DidPostClaude}{37.1}
\newcommand{\DidChangeClaude}{$-$30.0 [$-$40.7, $-$19.7]}

\newcommand{\DidChangeClaudeFourEight}{$-$11.9 [$-$21.9, $-$1.8]}

\newcommand{\PostCutOpenMCCmax}{0.130}

\newcommand{\PostCutChanceGapClaude}{16.0 [7.7, 24.5]}

\newcommand{\PostCutCthreeAMin}{70.2}
\newcommand{\PostCutCthreeAMax}{78.2}

\newcommand{\PostCutADRAllCISixOne}{33.8 [27.2, 40.5]}

\newcommand{\PostCutMaskDvsRandomSixOne}{$-$13.5 [$-$21.9, $-$5.2]}

\newcommand{\PostCutMaskDvsRandomClaude}{$-$13.5 [$-$24.0, $-$3.1]}

\newcommand{\PostCutMaskVvsNoneClaude}{32.6 [21.1, 44.2]}

\newcommand{\PostCutLFourADRClaude}{29.2}

\newcommand{\ReasonBudgetGptOss}{32,768}

\newcommand{\HcdNotAnalysed}{1}
\newcommand{\HcdAnalysed}{514}
\newcommand{\HcdChanged}{36}
\newcommand{\HcdUbsan}{227}
\newcommand{\HcdLinesRemoved}{150}
\newcommand{\HcdLines}{4,603}

\newcommand{\ToolASan}{284}
\newcommand{\ToolUBSan}{42}
\newcommand{\ToolMSan}{26}
\newcommand{\ToolLSan}{28}
\newcommand{\ToolTSan}{1}
\newcommand{\ToolAssert}{11}
\newcommand{\ToolTimeout}{8}
\newcommand{\ToolBehavior}{128}

\newcommand{\ToolTwo}{13}

\newcommand{\DirCRVulnLLM}{10.6 [5.0, 15.7]}

\newcommand{\DirCRQwenThirtyTwo}{5.4 [1.7, 8.9]}

\newcommand{\DirCRQwenThree}{5.2 [2.3, 8.3]}

\newcommand{\DirCRQwenThreeSix}{15.4 [10.0, 20.7]}

\newcommand{\DirCRGptOss}{13.5 [8.4, 18.6]}

\newcommand{\DirCRFiveSix}{22.7 [18.0, 27.6]}

\newcommand{\DirCRClaude}{65.7 [60.9, 70.4]}

\newcommand{\DirSigOpen}{5}
\newcommand{\DirOpenModels}{9}

\newcommand{\GuardDecFixMin}{12.0}
\newcommand{\GuardDecFixMax}{12.2}
\newcommand{\GuardDecAltMin}{3.2}
\newcommand{\GuardDecAltMax}{3.4}
\newcommand{\GuardDecDecoyMin}{2.1}
\newcommand{\GuardDecDecoyMax}{3.0}
\newcommand{\GuardDecVulnMin}{$-$11.8}
\newcommand{\GuardDecVulnMax}{$-$4.7}
\newcommand{\GuardDecFzeroModelMin}{98.2}

\newcommand{\KwGapClosedMin}{30.5}
\newcommand{\KwGapClosedMax}{52.9}
\newcommand{\PostCutKwGapClosedMin}{34.2}
\newcommand{\PostCutKwGapClosedMax}{46.7}
\newcommand{\BothVulnGptMin}{49.3}
\newcommand{\BothVulnGptMax}{54.9}
\newcommand{\TcRightClosedMin}{96.0}
\newcommand{\TcRightClosedMax}{99.0}
\newcommand{\RepSameClosedMin}{83.0}
\newcommand{\RepSameClosedMax}{86.9}
\newcommand{\CthreeAVzeroClosedMin}{67.0}
\newcommand{\CthreeAVzeroClosedMax}{87.2}
\newcommand{\CthreeAJointClosedMin}{69.3}
\newcommand{\CthreeAJointClosedMax}{86.1}
\newcommand{\CthreeADecoyFaultClosedMin}{70.0}
\newcommand{\CthreeADecoyFaultClosedMax}{95.0}

\newcommand{\TabPairs}{%
\multicolumn{10}{@{}l}{\emph{Commercial models}} \\
GPT-6.1-sol & 0.666 & 0.614 & 0.896 & 0.729 & 0.375 & 0.721 & 34.7 & 19.7 & 15.0 [11.0, 19.3] $\uparrow$ \\
GPT-5.6-sol & 0.613 & 0.589 & 0.753 & 0.661 & 0.236 & 0.624 & 26.0 & 23.1 & 2.9 [$-$1.2, 7.5] $\approx$ \\
Claude Opus 5.5 & 0.828 & 0.769 & 0.940 & 0.846 & 0.674 & 0.899 & 67.2 & 23.8 & 43.4 [39.2, 47.4] $\uparrow$ \\
Claude Opus 4.8 & 0.683 & 0.672 & 0.715 & 0.693 & 0.367 & 0.626 & 39.7 & 24.9 & 14.8 [10.5, 19.7] $\uparrow$ \\
\midrule
\multicolumn{10}{@{}l}{\emph{Open-weight models without a reasoning mode}} \\
Qwen2.5-Coder-32B & 0.527 & 0.533 & 0.439 & 0.481 & 0.055 & 0.518 & 10.2 & 24.2 & $-$14.0 [$-$16.8, $-$11.0] $\downarrow$ \\
CodeLlama-34B & 0.500 & 0.500 & 0.994 & 0.665 & 0.000 & 0.524 & 0.0 & 0.6 & $-$0.6 [$-$1.4, 0.0] $\approx$ \\
Qwen3-Coder-30B & 0.526 & 0.541 & 0.345 & 0.421 & 0.056 & 0.516 & 8.5 & 21.7 & $-$13.2 [$-$15.8, $-$10.3] $\downarrow$ \\
Qwen2.5-Coder-7B & 0.509 & 0.556 & 0.094 & 0.160 & 0.034 & 0.511 & 3.1 & 7.7 & $-$4.6 [$-$6.7, $-$2.6] $\downarrow$ \\
Qwen3-Coder-Next (8 bit) & 0.493 & 0.495 & 0.672 & 0.570 & $-$0.016 & 0.520 & 5.4 & 21.8 & $-$16.4 [$-$18.8, $-$14.0] $\downarrow$ \\
\midrule
\multicolumn{10}{@{}l}{\emph{Open-weight reasoning models}} \\
gpt-oss-20b & 0.568 & 0.566 & 0.580 & 0.573 & 0.135 & 0.511 & 21.6 & 25.0 & $-$3.4 [$-$7.3, 0.8] $\approx$ \\
VulnLLM-R-7B & 0.553 & 0.557 & 0.518 & 0.537 & 0.106 & 0.476 & 24.1 & 24.9 & $-$0.8 [$-$4.7, 3.0] $\approx$ \\
Qwen3.6-27B & 0.577 & 0.562 & 0.701 & 0.623 & 0.159 & 0.525 & 23.5 & 23.5 & 0.0 [$-$4.1, 4.0] $\approx$ \\
Foundation-Sec-8B-Reasoning & 0.511 & 0.507 & 0.832 & 0.630 & 0.030 & 0.503 & 16.0 & 14.7 & 1.3 [$-$1.6, 4.2] $\approx$ \\
\midrule
Surface-feature classifier$^{\dagger}$ & 0.523 & 0.526 & 0.462 & 0.492 & 0.046 & 0.530 & 5.8 & -- & -- \\
}

\newcommand{\TabKeyword}{%
\multicolumn{6}{@{}l}{\emph{Commercial models}} \\
GPT-6.1-sol & 67.2 & 27.2 & 13.6 & 40.0 [35.1, 44.8] & $\uparrow$ \\
GPT-5.6-sol & 58.6 & 25.7 & 13.2 & 32.9 [28.3, 37.4] & $\uparrow$ \\
Claude Opus 5.5 & 81.3 & 28.4 & 14.2 & 52.9 [48.3, 57.4] & $\uparrow$ \\
Claude Opus 4.8 & 55.9 & 25.4 & 12.4 & 30.5 [25.3, 35.6] & $\uparrow$ \\
\midrule
\multicolumn{6}{@{}l}{\emph{Open-weight models without a reasoning mode}} \\
Qwen2.5-Coder-32B & 11.9 & 6.6 & 2.4 & 5.2 [2.6, 7.6] & $\uparrow$ \\
CodeLlama-34B & 37.7 & 38.2 & 26.4 & $-$0.5 [$-$5.0, 4.2] & $\approx$ \\
Qwen3-Coder-30B & 8.3 & 4.8 & 1.9 & 3.5 [1.5, 5.7] & $\uparrow$ \\
Qwen2.5-Coder-7B & 9.0 & 8.3 & 3.2 & 0.7 [$-$1.8, 3.5] & $\approx$ \\
Qwen3-Coder-Next (8 bit) & 21.6 & 12.2 & 5.5 & 9.5 [5.9, 12.7] & $\uparrow$ \\
\midrule
\multicolumn{6}{@{}l}{\emph{Open-weight reasoning models}} \\
gpt-oss-20b & 39.7 & 22.8 & 11.7 & 16.9 [12.2, 21.3] & $\uparrow$ \\
VulnLLM-R-7B & 39.1 & 36.5 & 25.3 & 2.6 [$-$2.1, 7.0] & $\approx$ \\
Qwen3.6-27B & 38.3 & 17.3 & 8.9 & 21.0 [16.7, 25.3] & $\uparrow$ \\
Foundation-Sec-8B-Reasoning & 22.0 & 20.1 & 10.6 & 1.9 [$-$1.5, 5.7] & $\approx$ \\
}

\newcommand{\TabBlankGPT}{%
Nothing blanked & 34.7 & -- & 26.0 & -- \\
Dynamic slice blanked & 27.0 & $-$8.7 [$-$12.1, $-$5.4] $\downarrow$ & 23.7 & $-$9.8 [$-$13.4, $-$6.3] $\downarrow$ \\
DeepWukong slice blanked & 23.3 & $-$11.4 [$-$14.9, $-$8.0] $\downarrow$ & 18.3 & $-$7.7 [$-$11.0, $-$4.3] $\downarrow$ \\
Trigger line blanked & 30.1 & $-$4.6 [$-$7.5, $-$1.7] $\downarrow$ & 28.3 & 2.3 [$-$1.1, 5.6] $\approx$ \\
Random lines blanked & 35.8 & 1.0 [$-$1.7, 3.9] $\approx$ & 33.5 & 7.5 [4.4, 10.7] $\uparrow$ \\
Only the dynamic slice kept & 27.4 & $-$7.3 [$-$10.0, $-$4.4] $\downarrow$ & 23.1 & $-$2.9 [$-$6.4, 0.4] $\approx$ \\
}
\newcommand{\TabBlankHeadGPT}{%
 & \multicolumn{2}{c}{GPT-6.1-sol} & \multicolumn{2}{c}{GPT-5.6-sol} \\
\cmidrule(lr){2-3}\cmidrule(lr){4-5}
Condition & P-C & difference & P-C & difference \\
}
\newcommand{\TabBlankClaude}{%
Nothing blanked & 67.2 & -- & 39.7 & -- \\
Dynamic slice blanked & 65.7 & 0.2 [$-$2.5, 2.9] $\approx$ & 46.2 & $-$4.6 [$-$8.4, $-$0.8] $\downarrow$ \\
DeepWukong slice blanked & 62.8 & $-$4.4 [$-$7.2, $-$1.6] $\downarrow$ & 36.8 & $-$2.9 [$-$7.4, 1.5] $\approx$ \\
Trigger line blanked & 67.4 & 0.2 [$-$1.9, 2.3] $\approx$ & 47.6 & 7.9 [3.6, 11.8] $\uparrow$ \\
Random lines blanked & 65.5 & $-$1.7 [$-$3.9, 0.4] $\approx$ & 50.7 & 11.0 [7.6, 14.5] $\uparrow$ \\
Only the dynamic slice kept & 66.3 & $-$0.8 [$-$3.4, 1.7] $\approx$ & 36.6 & $-$3.1 [$-$8.4, 1.9] $\approx$ \\
}
\newcommand{\TabBlankHeadClaude}{%
 & \multicolumn{2}{c}{Claude Opus 5.5} & \multicolumn{2}{c}{Claude Opus 4.8} \\
\cmidrule(lr){2-3}\cmidrule(lr){4-5}
Condition & P-C & difference & P-C & difference \\
}

\newcommand{\TabBlankBegin}{%
\begin{tabular}{lrrlrrlrrlrrl}
}

\newcommand{\TabMark}{%
\multicolumn{7}{@{}l}{\emph{Commercial models}} \\
GPT-6.1-sol & 33.5 & 39.9 & 30.8 & 6.4 [2.9, 10.1] $\uparrow$ & 3.3 [$-$0.2, 6.8] $\approx$ & 4.2 [0.8, 7.8] $\uparrow$ \\
GPT-5.6-sol & 26.2 & 33.7 & 24.9 & 7.5 [3.6, 11.7] $\uparrow$ & 4.2 [0.0, 8.0] $\approx$ & 2.9 [$-$0.8, 6.7] $\approx$ \\
Claude Opus 5.5 & 68.2 & 71.7 & 65.7 & 3.5 [$-$0.2, 7.2] $\approx$ & 2.3 [$-$1.4, 5.9] $\approx$ & 0.6 [$-$3.5, 4.7] $\approx$ \\
Claude Opus 4.8 & 37.0 & 41.2 & 35.8 & 4.2 [$-$0.2, 8.3] $\approx$ & 1.5 [$-$2.7, 5.5] $\approx$ & 1.5 [$-$2.6, 5.5] $\approx$ \\
\midrule
\multicolumn{7}{@{}l}{\emph{Open-weight models without a reasoning mode}} \\
Qwen2.5-Coder-32B & 10.2 & 10.0 & 6.7 & $-$0.2 [$-$3.2, 3.2] $\approx$ & $-$1.2 [$-$4.5, 2.2] $\approx$ & 0.4 [$-$2.3, 3.0] $\approx$ \\
CodeLlama-34B & 0.0 & 0.2 & 0.0 & 0.2 [0.0, 0.6] $\approx$ & 0.0 [0.0, 0.0] $\approx$ & 0.2 [0.0, 0.6] $\approx$ \\
Qwen2.5-Coder-7B & 3.1 & 11.9 & 8.9 & 8.7 [5.5, 12.1] $\uparrow$ & 5.6 [2.6, 8.5] $\uparrow$ & $-$0.4 [$-$3.6, 2.7] $\approx$ \\
\midrule
\multicolumn{7}{@{}l}{\emph{Open-weight reasoning models}} \\
gpt-oss-20b & 23.9 & 22.5 & 20.4 & $-$1.5 [$-$5.5, 2.8] $\approx$ & $-$0.2 [$-$5.0, 4.2] $\approx$ & 1.9 [$-$2.1, 5.8] $\approx$ \\
VulnLLM-R-7B & 25.8 & 28.1 & 22.5 & 2.3 [$-$3.2, 7.8] $\approx$ & $-$1.2 [$-$6.7, 3.8] $\approx$ & 0.4 [$-$4.3, 5.2] $\approx$ \\
Qwen3.6-27B & 20.8 & 16.8 & 14.8 & $-$4.0 [$-$8.2, 0.0] $\approx$ & $-$2.9 [$-$7.5, 1.7] $\approx$ & 2.1 [$-$1.6, 5.8] $\approx$ \\
Foundation-Sec-8B-Reasoning & 16.2 & 12.5 & 12.9 & $-$3.7 [$-$8.2, 0.8] $\approx$ & $-$4.6 [$-$9.1, 0.0] $\approx$ & $-$2.5 [$-$6.6, 1.5] $\approx$ \\
}

\newcommand{\TabStrataBegin}{%
\begin{tabular}{lrrrrr}
}

\newcommand{\TabStatedJaccard}{%
\multicolumn{8}{@{}l}{\emph{Commercial models}} \\
GPT-6.1-sol & \textbf{70.2} (16.8) & \textbf{0.211} (0.046) & \textbf{0.299} (0.096) & \textbf{0.256} (0.098) & \textbf{0.303} (0.111) & \textbf{0.228} (0.109) & \textbf{0.273} (0.121) \\
GPT-5.6-sol & \textbf{66.0} (16.4) & \textbf{0.213} (0.046) & \textbf{0.276} (0.091) & \textbf{0.242} (0.095) & \textbf{0.284} (0.105) & \textbf{0.218} (0.104) & \textbf{0.260} (0.116) \\
Claude Opus 5.5 & \textbf{84.0} (17.3) & \textbf{0.226} (0.046) & \textbf{0.315} (0.097) & \textbf{0.274} (0.099) & \textbf{0.320} (0.114) & \textbf{0.241} (0.111) & \textbf{0.293} (0.124) \\
Claude Opus 4.8 & \textbf{60.9} (15.2) & \textbf{0.196} (0.046) & \textbf{0.278} (0.091) & \textbf{0.240} (0.094) & \textbf{0.283} (0.104) & \textbf{0.212} (0.103) & \textbf{0.249} (0.112) \\
\midrule
\multicolumn{8}{@{}l}{\emph{Open-weight models without a reasoning mode}} \\
Qwen2.5-Coder-32B & \textbf{11.7} (2.9) & \textbf{0.072} (0.016) & \textbf{0.059} (0.023) & \textbf{0.050} (0.023) & \textbf{0.058} (0.024) & \textbf{0.051} (0.024) & \textbf{0.052} (0.025) \\
CodeLlama-34B & \textbf{38.3} (26.1) & \textbf{0.099} (0.040) & \textbf{0.113} (0.093) & \textbf{0.116} (0.101) & \textbf{0.163} (0.125) & 0.099 (0.113) & \textbf{0.155} (0.137) \\
Qwen3-Coder-30B & \textbf{9.6} (2.4) & \textbf{0.031} (0.007) & \textbf{0.037} (0.015) & \textbf{0.038} (0.016) & \textbf{0.040} (0.016) & \textbf{0.033} (0.017) & \textbf{0.038} (0.018) \\
Qwen2.5-Coder-7B & \textbf{10.4} (3.7) & \textbf{0.077} (0.027) & \textbf{0.062} (0.032) & \textbf{0.052} (0.033) & \textbf{0.065} (0.034) & \textbf{0.052} (0.033) & \textbf{0.057} (0.034) \\
Qwen3-Coder-Next (8 bit) & \textbf{21.3} (6.1) & \textbf{0.069} (0.019) & \textbf{0.094} (0.039) & \textbf{0.088} (0.040) & \textbf{0.103} (0.045) & \textbf{0.082} (0.045) & \textbf{0.098} (0.048) \\
\midrule
\multicolumn{8}{@{}l}{\emph{Open-weight reasoning models}} \\
gpt-oss-20b & \textbf{44.1} (14.3) & \textbf{0.158} (0.043) & \textbf{0.235} (0.082) & \textbf{0.209} (0.085) & \textbf{0.245} (0.097) & \textbf{0.191} (0.095) & \textbf{0.227} (0.105) \\
VulnLLM-R-7B & \textbf{42.0} (27.1) & \textbf{0.094} (0.045) & \textbf{0.158} (0.095) & \textbf{0.145} (0.102) & \textbf{0.200} (0.126) & \textbf{0.133} (0.117) & \textbf{0.191} (0.141) \\
Qwen3.6-27B & \textbf{42.3} (11.1) & \textbf{0.176} (0.041) & \textbf{0.207} (0.068) & \textbf{0.174} (0.072) & \textbf{0.200} (0.074) & \textbf{0.161} (0.077) & \textbf{0.184} (0.082) \\
Foundation-Sec-8B-Reasoning & \textbf{24.2} (12.5) & \textbf{0.106} (0.043) & \textbf{0.131} (0.073) & \textbf{0.117} (0.075) & \textbf{0.148} (0.084) & \textbf{0.107} (0.083) & \textbf{0.136} (0.090) \\
}

\newcommand{\TabConditionsBegin}{%
\begin{tabular}{lccccccccccc}
}

\newcommand{\TabCWEBegin}{%
\begin{tabular}{llrrrrrrrrrrrr}
}

\newcommand{\ExampleRead}{%
1 & \texttt{seamless\_process(STREAM s)} &  &  &  & 1 &  &  \\
2 & \hspace*{0.5em}\texttt{\{} &  &  &  &  &  &  \\
3 & \hspace*{2.5em}\texttt{unsigned int pkglen;} &  &  &  &  &  & 3 \\
4 & \hspace*{2.5em}\texttt{char *buf;} &  &  &  &  &  &  \\
 & \textcolor{gray}{\hspace*{2.0em}\texttt{struct stream packet = *s;}} &  &  &  &  & \textcolor{gray}{5} &  \\
 & \textcolor{gray}{\hspace*{2.0em}\texttt{if (!s\_check(s))}} &  &  &  &  & \textcolor{gray}{7} &  \\
 & \textcolor{gray}{\hspace*{2.0em}\texttt{\{}} &  &  &  &  & \textcolor{gray}{8} &  \\
 & \textcolor{gray}{\hspace*{4.0em}\texttt{rdp\_protocol\_error("seamless\_process(), stream is in unstable state", \&packet);}} &  &  &  &  & \textcolor{gray}{9} &  \\
 & \textcolor{gray}{\hspace*{2.0em}\texttt{\}}} &  &  &  &  & \textcolor{gray}{10} &  \\
5 & \hspace*{0.5em}\texttt{} &  &  &  &  &  &  \\
6 & \hspace*{2.5em}\texttt{pkglen = s{-}{>}end {-} s{-}{>}p;} & 6 & 6 & 6 & 6 &  & 6 \\
7 & \hspace*{2.5em}\texttt{/* str\_handle\_lines requires null terminated strings */} &  &  &  &  &  &  \\
8 & \hspace*{2.0em}\texttt{buf = xmalloc(pkglen + 1);} & 8 & 8 & 8 & 8 &  & 8 \\
\textbf{9}$\blacktriangleright$ & \hspace*{2.0em}\texttt{STRNCPY(buf, (char *) s{-}{>}p, pkglen + 1);} & 9 & 9 & 9 & 9 &  & 9 \\
10 & \hspace*{2.0em}\texttt{str\_handle\_lines(buf, \&seamless\_rest, seamless\_line\_handler, NULL);} &  &  &  &  &  & 10 \\
11 & \texttt{} &  &  &  &  &  &  \\
12 & \hspace*{2.0em}\texttt{xfree(buf);} &  &  &  &  &  &  \\
13 & \texttt{\}} &  &  &  &  &  &  \\
}

\newcommand{\ExampleStaleShort}{%
1 & \texttt{static char ** split(const char *arg, const char *delim) \{} &  & 1 &  & 1 &  &  \\
2 & \hspace*{1.0em}\texttt{char *copy = dupstr(arg);} & 2 & 2 & 2 & 2 &  &  \\
3 & \hspace*{1.0em}\texttt{char **result = NULL;} & 3 & 3 & 3 & 3 &  &  \\
4 & \hspace*{1.0em}\texttt{int i = 0;} & 4 & 4 & 4 & 4 &  &  \\
5 & \texttt{} &  &  &  &  &  &  \\
6 & \hspace*{1.0em}\texttt{for (char *cptr = strtok(copy, delim); cptr; cptr = strtok(NULL, delim)) \{} & 6 & 6 & 6 & 6 &  &  \\
7 & \hspace*{2.0em}\texttt{char **tmp = realloc (result, sizeof *result * (i + 1));} & 7 & 7 & 7 & 7 &  & 7 \\
8 & \hspace*{2.0em}\texttt{if (!tmp \&\& result) \{} & 8 & 8 & 8 & 8 &  & 8 \\
9--15 & \hspace*{1.0em}\textit{error branch of the first realloc, which frees the tokens, result, and copy} &  &  &  &  &  &  \\
16 & \hspace*{2.0em}\texttt{result = tmp;} &  & 16 & 16 & 16 &  & 16 \\
17 & \hspace*{2.0em}\texttt{result[i++] = dupstr(cptr);} & 17 & 17 & 17 & 17 &  & 17 \\
18 & \hspace*{1.0em}\texttt{\}} & 18 &  &  &  &  &  \\
19 & \texttt{} &  &  &  &  &  &  \\
20 & \hspace*{1.0em}\texttt{free(copy);} &  & 20 &  & 20 &  & 20 \\
21 & \texttt{} &  &  &  &  &  &  \\
22 & \hspace*{1.0em}\texttt{if (i) \{} & 22 &  & 22 & 22 &  &  \\
23 & \hspace*{2.0em}\texttt{char **tmp = realloc(result, sizeof *result * (i + 1));} & 23 & 23 & 23 & 23 &  & 23 \\
24 & \hspace*{2.0em}\texttt{if (!tmp) \{} & 24 & 24 & 24 & 24 &  & 24 \\
25 & \hspace*{3.0em}\texttt{while (i {>} 0) \{} & 25 &  &  & 25 &  &  \\
26 & \hspace*{2.0em}\texttt{free(result[{-}{-}i]);} & 26 &  &  & 26 &  &  \\
27 & \hspace*{3.0em}\texttt{\}} &  &  &  &  &  &  \\
28 & \hspace*{3.0em}\texttt{free(result);} &  &  &  & 28 &  &  \\
\textbf{29}$\blacktriangleright$ & \hspace*{3.0em}\texttt{free(copy);} & 29 & 29 & 29 & 29 & 29 & 29 \\
30 & \hspace*{3.0em}\texttt{return NULL;} &  &  &  &  &  &  \\
31 & \hspace*{2.0em}\texttt{\}} &  &  &  &  &  &  \\
32 & \hspace*{2.0em}\texttt{result = tmp;} &  &  &  &  &  &  \\
33 & \hspace*{2.0em}\texttt{result[i++] = NULL;} &  &  &  &  &  &  \\
34 & \hspace*{1.0em}\texttt{\}} &  &  &  &  &  &  \\
35 & \texttt{} &  &  &  &  &  &  \\
36 & \hspace*{1.0em}\texttt{return result;} &  &  &  &  &  &  \\
37 & \texttt{\}} &  &  &  &  &  &  \\
}
\newcommand{\TabStaticDyn}{%
Joern & 79.5 [77.2, 81.7] & 32.4 & 1.14$\times$ & 0.545 & 39.8 \\
dg & 97.4 [96.3, 98.3] & 39.0 & 1.56$\times$ & 0.600 & 89.1 \\
DeepWukong, backward to the trigger line & 85.0 [82.8, 87.0] & 42.9 & 1.50$\times$ & 0.500 & 53.6 \\
}
\newcommand{\TabCthreeAHead}{%
GPT-6.1-sol & GPT-5.6-sol & Claude Opus 5.5 & Claude Opus 4.8
}
\newcommand{\TabCthreeABegin}{%
\begin{tabular}{lcccc}
}
\newcommand{\TabCthreeA}{%
All versions & 92.9 [89.7, 95.7] & 91.2 [88.1, 94.0] & 94.6 [91.7, 97.0] & 86.3 [82.4, 90.0] \\
Function body hidden (shortcut) & 51.6 [49.6, 53.6] & 49.0 [47.0, 51.0] & 46.5 [44.6, 48.4] & 51.8 [49.9, 53.9] \\
Answer \emph{fault} for every version (shortcut) & 42.2 & 42.2 & 42.2 & 42.2 \\
Vulnerable version & 84.4 [77.1, 90.8] & 78.9 [70.6, 86.2] & 87.2 [80.7, 92.7] & 67.0 [57.8, 75.2] \\
Fixed version & 99.1 [97.2, 100.0] & 98.2 [95.4, 100.0] & 98.2 [95.4, 100.0] & 100.0 [96.6, 100.0] \\
Both real versions of a function & 83.5 [76.1, 89.9] & 77.1 [68.8, 84.4] & 85.3 [78.0, 91.7] & 67.0 [57.8, 75.2] \\
Alternative fix & 100.0 [96.0, 100.0] & 100.0 [96.0, 100.0] & 100.0 [96.0, 100.0] & 98.9 [96.7, 100.0] \\
Decoy that still faults & 85.0 [70.0, 100.0] & 90.0 [75.0, 100.0] & 95.0 [85.0, 100.0] & 70.0 [50.0, 90.0] \\
Decoy that removed the fault & 100.0 [75.8, 100.0] & 100.0 [75.8, 100.0] & 100.0 [75.8, 100.0] & 100.0 [75.8, 100.0] \\
Renaming keeps the answer & 99.5 [98.4, 100.0] & 96.3 [93.2, 98.4] & 98.9 [97.4, 100.0] & 96.8 [94.2, 98.9] \\
Fault predicted and line within two lines (all faulting versions) & 81.2 [73.1, 88.8] & 77.2 [69.2, 85.0] & 86.1 [79.1, 92.4] & 69.3 [59.9, 78.2] \\
Line within two lines, versions predicted to fault & 97.0 [92.4, 100.0] & 96.9 [91.9, 100.0] & 98.9 [96.5, 100.0] & 100.0 [94.9, 100.0] \\
\quad random line (shortcut) & 16.4 & 15.9 & 15.9 & 14.0 \\
\quad line with most risky tokens (shortcut) & 26.6 [17.1, 36.1] & 26.7 [17.3, 36.3] & 27.3 [17.9, 36.4] & 29.3 [19.6, 40.1] \\
}

\newcommand{\TabLfourHead}{%
Version shown to the model & Correct verdict & GPT-6.1-sol & GPT-5.6-sol & Claude Opus 5.5 & Claude Opus 4.8
}
\newcommand{\TabLfourBegin}{%
\begin{tabular}{llcccc}
}

\newcommand{\TabStrataEvidenceBegin}{%
\begin{tabular}{@{}lllll@{}}
}

\newcommand{\FigFindings}{%
\begin{tikzpicture}
\begin{groupplot}[group style={group size=4 by 1, horizontal sep=2.5mm}, width=0.305\linewidth, height=40mm, ymin=0.4, ymax=4.6, ytick={1,...,4}, y dir=normal, tick label style={font=\scriptsize}, title style={font=\scriptsize, align=center}, xlabel style={font=\scriptsize}, xlabel={points}, xmajorgrids, grid style={black!10},]
\nextgroupplot[title={L1: P-C\\minus chance}, yticklabels={Claude Opus 4.8,Claude Opus 5.5,GPT-5.6-sol,GPT-6.1-sol},]
\addplot[only marks, mark=*, mark size=1.6pt, color=vuBlue, error bars/.cd, x dir=both, x explicit, error bar style={vuBlue, thick}] coordinates {(15.0,4.17) += (4.3,0) -= (4.0,0) (2.9,3.17) += (4.6,0) -= (4.1,0) (43.4,2.17) += (4.0,0) -= (4.2,0) (14.8,1.17) += (4.9,0) -= (4.3,0)};\label{fmain}
\addplot[only marks, mark=square*, mark size=1.5pt, color=vuRed, error bars/.cd, x dir=both, x explicit, error bar style={vuRed, thick}] coordinates {(6.8,3.83) += (9.2,0) -= (7.9,0) (0.9,2.83) += (8.8,0) -= (8.1,0) (16.0,1.83) += (8.5,0) -= (8.3,0) (6.3,0.83) += (8.8,0) -= (8.4,0)};\label{fpost}
\draw[black!45, densely dotted] (axis cs:0,0.4) -- (axis cs:0,4.6);
\nextgroupplot[title={L2: trigger line cited,\\minus keyword rule}, yticklabels={}, xmin=0,]
\addplot[only marks, mark=*, mark size=1.6pt, color=vuBlue, error bars/.cd, x dir=both, x explicit, error bar style={vuBlue, thick}] coordinates {(40.0,4.17) += (4.8,0) -= (4.9,0) (32.9,3.17) += (4.5,0) -= (4.6,0) (52.9,2.17) += (4.5,0) -= (4.6,0) (30.5,1.17) += (5.1,0) -= (5.2,0)};
\addplot[only marks, mark=square*, mark size=1.5pt, color=vuRed, error bars/.cd, x dir=both, x explicit, error bar style={vuRed, thick}] coordinates {(46.7,3.83) += (11.5,0) -= (10.8,0) (34.2,2.83) += (10.1,0) -= (10.5,0) (41.5,1.83) += (10.5,0) -= (10.4,0) (37.2,0.83) += (10.6,0) -= (10.1,0)};
\draw[black!45, densely dotted] (axis cs:0,0.4) -- (axis cs:0,4.6);
\nextgroupplot[title={L3: blank $D$,\\minus blank random}, yticklabels={},]
\addplot[only marks, mark=*, mark size=1.6pt, color=vuBlue, error bars/.cd, x dir=both, x explicit, error bar style={vuBlue, thick}] coordinates {(-8.7,4.17) += (3.3,0) -= (3.4,0) (-9.8,3.17) += (3.5,0) -= (3.6,0) (0.2,2.17) += (2.7,0) -= (2.7,0) (-4.6,1.17) += (3.8,0) -= (3.8,0)};
\addplot[only marks, mark=square*, mark size=1.5pt, color=vuRed, error bars/.cd, x dir=both, x explicit, error bar style={vuRed, thick}] coordinates {(-13.5,3.83) += (8.3,0) -= (8.4,0) (-7.3,2.83) += (8.3,0) -= (7.3,0) (-13.5,1.83) += (10.4,0) -= (10.5,0) (-6.2,0.83) += (8.3,0) -= (8.4,0)};
\draw[black!45, densely dotted] (axis cs:0,0.4) -- (axis cs:0,4.6);
\nextgroupplot[title={L5: outcome accuracy,\\minus guard rule}, yticklabels={},]
\addplot[only marks, mark=*, mark size=1.6pt, color=vuBlue, error bars/.cd, x dir=both, x explicit, error bar style={vuBlue, thick}] coordinates {(12.6,4.17) += (5.1,0) -= (5.1,0) (10.9,3.17) += (4.8,0) -= (4.7,0) (14.3,2.17) += (4.9,0) -= (4.9,0) (6.0,1.17) += (4.7,0) -= (4.9,0)};
\addplot[only marks, mark=square*, mark size=1.5pt, color=vuRed, error bars/.cd, x dir=both, x explicit, error bar style={vuRed, thick}] coordinates {(-0.4,3.83) += (8.0,0) -= (8.0,0) (-2.8,2.83) += (8.3,0) -= (8.1,0) (2.0,1.83) += (7.8,0) -= (8.3,0) (-6.0,0.83) += (8.3,0) -= (8.0,0)};
\draw[black!45, densely dotted] (axis cs:0,0.4) -- (axis cs:0,4.6);
\end{groupplot}
\node[anchor=north, font=\scriptsize] at ([yshift=-8mm]group c2r1.south east) {\ref{fmain} pre-cutoff set (CVEs up to 2022)\quad \ref{fpost} CVEs published after the training cutoffs};
\end{tikzpicture}
}
\newcommand{\FigPairOutcomes}{%
\begin{tikzpicture}
\begin{axis}[xbar stacked, bar width=3.2mm, width=0.70\linewidth, height=73mm, xmin=0, xmax=100, ymin=-13.9, ymax=0.7, ytick={0.0,-1.0,-2.0,-3.0,-4.6,-5.6,-6.6,-7.6,-8.6,-10.2,-11.2,-12.2,-13.2}, yticklabels={GPT-6.1-sol,GPT-5.6-sol,Claude Opus 5.5,Claude Opus 4.8,Qwen2.5-Coder-32B,CodeLlama-34B,Qwen3-Coder-30B,Qwen2.5-Coder-7B,Qwen3-Coder-Next (8 bit),gpt-oss-20b,VulnLLM-R-7B,Qwen3.6-27B,Foundation-Sec-8B-Reasoning}, y dir=normal, tick label style={font=\scriptsize}, xlabel={share of matched pairs (\%)}, xlabel style={font=\scriptsize}, legend style={font=\scriptsize, at={(0.5,1.02)}, anchor=south, legend columns=2, draw=none, column sep=3mm}, legend image code/.code={\draw[##1] (0cm,-0.1cm) rectangle (0.3cm,0.1cm);}, xmajorgrids, grid style={black!10}]
\addplot[fill=vuBlue, draw=vuBlue] coordinates {(34.7,0.0) (26.0,-1.0) (67.2,-2.0) (39.7,-3.0) (10.2,-4.6) (0.0,-5.6) (8.5,-6.6) (3.1,-7.6) (5.4,-8.6) (21.6,-10.2) (24.1,-11.2) (23.5,-12.2) (16.0,-13.2)};
\addlegendentry{pair-wise correct (P-C)}
\addplot[fill=black!70, draw=black!70] coordinates {(1.5,0.0) (3.3,-1.0) (1.5,-2.0) (3.1,-3.0) (4.8,-4.6) (0.0,-5.6) (3.3,-6.6) (1.2,-7.6) (6.9,-8.6) (8.1,-10.2) (13.5,-11.2) (8.1,-12.2) (13.7,-13.2)};
\addlegendentry{pair-wise reversed (P-R)}
\addplot[fill=black!20, draw=black!30] coordinates {(8.9,0.0) (21.4,-1.0) (4.6,-2.0) (25.4,-3.0) (51.4,-4.6) (0.6,-5.6) (62.2,-6.6) (89.4,-7.6) (26.0,-8.6) (33.9,-10.2) (34.7,-11.2) (21.8,-12.2) (3.1,-13.2)};
\addlegendentry{both judged benign (P-B)}
\addplot[fill=vuRed!85, draw=vuRed] coordinates {(54.9,0.0) (49.3,-1.0) (26.8,-2.0) (31.8,-3.0) (33.7,-4.6) (99.4,-5.6) (26.0,-6.6) (6.2,-7.6) (61.7,-8.6) (36.4,-10.2) (27.7,-11.2) (46.6,-12.2) (67.2,-13.2)};
\addlegendentry{both judged vulnerable (P-V)}
\end{axis}
\end{tikzpicture}
}

\newcommand{\parag}[1]{\par\vspace{3pt}\noindent\textbf{#1}\ }

\theoremstyle{definition}
\newtheorem{caseexample}{Example}
\theoremstyle{plain}
\newtheorem{ownproposition}{Proposition}     
\newcommand{\LOname}{line Jaccard}
\newcommand{\LOName}{Line Jaccard}
\newcommand{\LOsym}{J}
\newcommand{\LOhead}{\shortstack[r]{Line\\Jaccard}}
\newcommand{\LOOname}{leave-one-out}

\newcommand{\vv}{f_{v}}
\newcommand{\ff}{f_{f}}

\title{Right Verdict, Right Reason? An Execution-Grounded Study of Whether LLM Vulnerability Detectors Understand Vulnerabilities}
\titlerunning{Right Verdict, Right Reason?}

\author{Leizhen Zhang}
{School of Computing and Informatics, University of Louisiana at Lafayette, Lafayette, LA, USA}
{leizhen.zhang1@louisiana.edu}
{}{}

\author{Sheng Chen\footnote{Corresponding author.}}
{School of Computing and Informatics, University of Louisiana at Lafayette, Lafayette, LA, USA}
{sheng.chen@louisiana.edu}
{}{}

\authorrunning{L. Zhang and S. Chen}

\Copyright{Leizhen Zhang and Sheng Chen}

\ccsdesc[500]{Security and privacy~Software security engineering}
\ccsdesc[500]{Software and its engineering~Software testing and debugging}
\ccsdesc[300]{Computing methodologies~Machine learning}

\keywords{vulnerability detection, large language models, dynamic program slicing, explanation evaluation, counterfactual evaluation}

\category{}
\relatedversion{}

\EventEditors{}
\EventNoEds{0}
\EventLongTitle{}
\EventShortTitle{ECOOP 2027}
\EventAcronym{ECOOP}
\EventYear{2027}
\EventDate{}
\EventLocation{}
\EventLogo{}
\SeriesVolume{}
\ArticleNo{}

\begin{document}
\raggedbottom   
\setlength{\emergencystretch}{1em}   
\maketitle

\begin{abstract}
Large language models (LLMs) are increasingly used to detect vulnerabilities, and their benchmark scores are often read as evidence that they understand the vulnerabilities they report. We test this reading against program execution. For 515 C and C++ functions from BigVul and PrimeVul whose vulnerability we reproduce with a proof of concept, we identify the line at which the fault occurs and the statements on which the faulting run depended. These facts ground a ladder of five tests, from telling a vulnerable function from its fixed version, through locating the fault and reacting to its removal, to judging execution-confirmed edits and predicting the outcome of a concrete run. Each test is paired with a baseline that needs no understanding, such as a rule that picks lines with risky calls, and we repeat the study on 195 vulnerabilities published after the training cutoffs of the models. Four findings emerge. First, telling a vulnerable function from its fixed version is hard. The open-weight models separate no more pairs than chance, although five of them show a weak sensitivity to the fix, and the commercial models still label many screened fixed versions vulnerable. Second, the commercial models locate the fault beyond a keyword rule, while removing that evidence changes the verdicts of the strongest model only on the newer vulnerabilities. Third, given the program input, the commercial models predict whether a run faults better than a rule that looks for added checks, and the prediction follows the input. Fourth, on the newer vulnerabilities this advantage disappears and the lead of the strongest model narrows, while localization persists. We release the framework, the per-function evidence, and an execution-confirmed counterfactual benchmark.
\end{abstract}

\section{Introduction}
\label{sec:intro}

Learning-based vulnerability detection has progressed from code gadgets and program graphs~\cite{li2018vuldeepecker,zhou2019devign,Reveal} to fine-tuned transformers~\cite{Fu:2022:LineVul,steenhoek2024dataflow}, and general-purpose large language models (LLMs) now return both a verdict and a natural-language explanation for a function~\cite{khare2023understanding,ullah2024llms,steenhoek2024toerr,ding2024vulnerability}. Developers use these explanations to decide which lines to inspect and whether a patch addresses the fault. Label noise and training leakage inflate reported scores~\cite{croft2023dataquality,ding2024vulnerability}, detectors exploit artifacts of data collection~\cite{Broken:2023:Imgrund,arp2022dos}, and they struggle to tell a vulnerable function from its fixed version~\cite{risse2024uncovering,rabin2021generalizability}. A correct verdict can therefore rest on the wrong reason. We ask \emph{when an LLM detector is right, is it right for the reason that makes the function vulnerable?}

Answering this question requires three ingredients.

\parag{C1. A reference for the reason.} Existing studies compare explanations with patch lines, static slices, or human judgment~\cite{ganz2021explaining,hu2023interpreters,cheng2024beyond,ullah2024llms}. Patch lines describe the repair and may miss the fault site, static slices depend on the analysis and the slicing criterion~\cite{Cheng:2021:DeepWuKong,Joern}, and human rationales are costly and subjective. We replay a proof of concept for every function and take two facts of the faulting run, the \emph{trigger line} at which the fault occurs and the \emph{dynamic dependence slice} of the statements on which the run depended~\cite{korel1988dynamic,agrawal1990dynamic}. The slice covers \SliceSizeDyn\% of a function on average and needs no manual annotation.

\parag{C2. Baselines without understanding.} Correct verdicts can follow from memorized Common Vulnerabilities and Exposures (CVE) entries~\cite{carlini2021extracting,Yang2024}, from surface features such as an added bounds check, or from a bias toward \emph{vulnerable}, and plausible citations can follow from risky keywords. Every test is therefore paired with a shortcut that needs no understanding or with a control that applies a comparable change without targeting the evidence.

\parag{C3. Interventions confirmed by execution.} A model can cite the correct lines without using them~\cite{jacovi2020faithfully,turpin2023unfaithful}. We blank or mark the evidence in the prompt, and we build edited versions of each function whose labels we establish by running the proof of concept on them.

\begin{figure}[t]
\centering
\begin{tikzpicture}[node distance=2.5mm and 4mm, every node/.style={font=\scriptsize},
  box/.style={draw, rounded corners=2pt, align=center, text width=31mm, minimum height=8.5mm, inner sep=2pt},
  res/.style={box, thick, fill=black!6},
  lvl/.style={draw, rounded corners=2pt, align=left, text width=52mm, minimum height=7mm, inner sep=2pt},
  arr/.style={-{Stealth[length=1.8mm]}, semithick}]
\node[box] (data) {BigVul and PrimeVul:\\function $\vv$, fixed version $\ff$};
\node[box, below=of data] (poc) {proof of concept $(H,x)$\\replayed on both builds};
\node[box, below=of poc] (screen) {fixed version screened\\manual review, dynamic tests};
\node[res, right=of data] (trig) {trigger line $t$\\where the fault occurs};
\node[res, below=of trig] (dyn) {dynamic slice $D$\\what the run depended on};
\node[res, below=of dyn] (var) {variants of $\vv$\\labeled by execution};
\node[lvl, right=6mm of trig.north east, anchor=north west] (l1) {\textbf{L1 / RQ1} verdict on $\vv$ and $\ff$\\\textit{vs.\ surface classifier, chance $v(1-v)$}};
\node[lvl, below=1.2mm of l1] (l2) {\textbf{L2 / RQ2} cited and attributed lines\\\textit{vs.\ keyword rule, random lines}};
\node[lvl, below=1.2mm of l2] (l3) {\textbf{L3 / RQ3} verdict with evidence blanked or marked\\\textit{vs.\ as many random lines}};
\node[lvl, below=1.2mm of l3] (l4) {\textbf{L4 / RQ4} verdict on confirmed edits\\\textit{vs.\ renaming, neutral edits, decoys}};
\node[lvl, below=1.2mm of l4] (l5) {\textbf{L5 / RQ5} outcome of one run, given the input\\\textit{vs.\ guard rules, benign input, hidden body}};
\draw[arr] (data) -- (poc); \draw[arr] (poc) -- (screen);
\draw[arr] (poc.east) -- (trig.west); \draw[arr] (poc.east) -- (dyn.west); \draw[arr] (screen.east) -- (var.west);
\draw[arr] (trig.east) -- (l2.west); \draw[arr] (dyn.east) -- (l2.west); \draw[arr] (dyn.east) -- (l3.west);
\draw[arr] (var.east) -- (l4.west); \draw[arr] (var.east) -- (l5.west);
\node[draw, dashed, rounded corners=2pt, fit=(l1)(l5), inner sep=1.5mm] (ladder) {};
\end{tikzpicture}
\caption{Overview of the study. Every function comes with a replayed proof of concept, from which we take the execution-grounded reference (Section~\ref{sec:reference}) and the labels of edited variants. Each of the five tests answers one research question and is read against its own shortcut or control (Section~\ref{sec:design}). Thirteen models are tested on \NumFunctions{} functions, and the tests are repeated on \PostCutStudyCVEs{} CVEs published after the training cutoffs (Section~\ref{sec:postcut}).}
\label{fig:overview}
\end{figure}

\parag{Study.} These ingredients define a ladder of five tests, and each research question addresses one level (Figure~\ref{fig:overview} and Table~\ref{tab:levels}). \textbf{RQ1} asks whether a model separates a vulnerable function from its fixed version (L1), \textbf{RQ2} whether its evidence points to the fault (L2), \textbf{RQ3} whether its verdict depends on that evidence (L3), \textbf{RQ4} whether its verdict follows execution-confirmed edits of the function (L4), and \textbf{RQ5} whether it predicts the outcome of one concrete run (L5). We evaluate nine open-weight and four commercial models on \NumFunctions{} C and C++ functions from BigVul and PrimeVul~\cite{Fan2020,ding2024vulnerability} and repeat the experiments on \PostCutStudyCVEs{} CVEs that were published and fixed after the commercial models' training cutoffs. Four findings emerge.

\begin{itemize}
\item \textbf{F1. Telling a vulnerable function from its fix is hard (RQ1).} No open-weight model separates more pairs than chance. Five of them show a weak sensitivity to the fix, since more of their pairs with differing verdicts are correct than reversed, yet they mostly give both versions the same verdict. The commercial models do better, but the two GPT models label both versions vulnerable for about half of the pairs, although every fixed version passed our dynamic and manual screening.
\item \textbf{F2. Localization exceeds the shortcuts, and reliance on it is partial (RQ2, RQ3).} The commercial models cite the trigger line far more often than a keyword rule that cites as many lines, and their cited lines match the dynamic slice better than two of three static slices. Blanking the dynamic slice lowers the pair discrimination of three of them. The verdicts of Claude Opus 5.5 persist.
\item \textbf{F3. Verdicts withstand label-preserving edits, and a concrete input enables execution reasoning (RQ4, RQ5).} Renaming and neutral edits leave the verdicts unchanged. With the driver and the input in view, all four commercial models predict whether a run faults better than a rule that knows the vulnerable version and accepts every added check, mainly because they predict the developer's fixes clean, and their prediction follows the input.
\item \textbf{F4. Part of the strongest results depends on known vulnerabilities (Section~\ref{sec:rq4}).} On the CVEs published after the cutoffs, localization persists, the advantage over the guard rule at L5 disappears, and the lead of Claude Opus 5.5 at L1 narrows, also relative to the most recent CVEs of the pre-cutoff set.
\end{itemize}

\parag{Contributions.} (1) An execution-grounded reference for vulnerability explanations, built from replayed proofs of concept and compared with patch lines and three static slicers (Section~\ref{sec:reference}). (2) A five-level test ladder in which every level has a shortcut or a control, and an execution-confirmed counterfactual benchmark of \VariantsConfirmed{} variants (Sections~\ref{sec:design} and~\ref{sec:benchmark}). (3) An evaluation of thirteen models before and after their training cutoffs that separates localization, reliance on evidence, and execution reasoning (Sections~\ref{sec:rq1} to~\ref{sec:rq4}). We release the framework, the per-function evidence, and the benchmark.

\begin{table}[t]
\centering
\caption{The five tests. Each level answers one research question. A shortcut needs no understanding of the vulnerability, and a control applies a comparable change without targeting the evidence. The labels of L4 and L5 are confirmed by execution.}
\label{tab:levels}
\footnotesize
\setlength{\tabcolsep}{4pt}
\begin{tabular}{@{}l>{\raggedright\arraybackslash}p{0.27\linewidth}>{\raggedright\arraybackslash}p{0.39\linewidth}@{}}
\toprule
RQ and level & Test & Compared with (shortcut or control) \\
\midrule
\textbf{RQ1} L1 Discriminate & Verdict on $\vv$ and on $\ff$, each shown alone & shortcut: a classifier on surface features such as length; control: a detector that ignores its input, with chance $v(1-v)$ \\
\textbf{RQ2} L2 Localize & Cited or top-ranked lines against the trigger line and the slices & shortcuts: a rule that cites lines with risky calls, and random lines \\
\textbf{RQ3} L3 Evidence sensitivity & Verdict after the evidence is blanked or marked in the prompt & control: as many random lines blanked or marked \\
\textbf{RQ4} L4 Counterfactual & Verdict on execution-confirmed edits of the function & controls: renaming and neutral edits, which keep the label; shortcut: accepting every added check, which decoys defeat \\
\textbf{RQ5} L5 Execution & Outcome and faulting line of one run, given driver and input & shortcuts: rules that look for bounds checks; controls: the same driver with a benign input, and the function body hidden \\
\bottomrule
\end{tabular}
\end{table}

\section{Execution-Grounded Reference}
\label{sec:reference}

Judging whether a detector is right for the right reason requires an account of that reason that is independent of the detector, of the patch, and of any particular analysis tool. This section defines such an account from a single confirmed faulting execution, describes how we obtain it, illustrates it on two functions, and compares it with the references of prior work.

\subsection{Definition and Scope}
\label{sec:instances}

\begin{definition}[Instance]
\label{def:instance}
An \emph{instance} is a tuple $(\vv, \ff, H, x)$. $\vv$ is a function labeled vulnerable, $\ff$ is its version after the fixing commit, $H$ is a harness that builds an executable program $H[f]$ around a version $f$ of the function, and $x$ is an input. The instance is \emph{confirmed} when the run $H[\vv](x)$ exhibits a fault that an oracle detects and the run $H[\ff](x)$ is free of that fault.
\end{definition}

The oracle is a sanitizer report~\cite{serebryany2012asan,stepanov2015msan}, a failed assertion, or a behavioral difference in the exit code. We write $\mathit{lines}(\vv)$ for the non-empty source lines of $\vv$. All references defined below are subsets of $\mathit{lines}(\vv)$, so that they can be compared with the line numbers a detector reports for $\vv$.

\begin{definition}[Trigger line]
\label{def:trigger}
Let $\rho$ be the faulting run $H[\vv](x)$. The \emph{trigger line} $t\in\mathit{lines}(\vv)$ is the line of the innermost stack frame of the fault report that lies inside $\vv$.
\end{definition}

When the fault is raised by a statement of $\vv$ itself, $t$ is the faulting statement, and when it is raised inside a callee such as \texttt{memcpy}, $t$ is the call in $\vv$ that was executing. Some confirmed instances have no report frame inside $\vv$, because the two runs differ only in their exit status, as for denial of service or missing authorization checks, or because the report frames lie outside $\vv$. For these we locate where the fix decides the outcome.

\begin{definition}[Patch-derived trigger line]
\label{def:patchtrigger}
Let $C\subseteq\mathit{lines}(\vv)$ be the lines on which the fix acts: every line that the fix replaces or deletes, and, for every insertion, the first statement of $\vv$ after the insertion point. The \emph{patch-derived trigger line} is the first line of $C$ that is executed in $\rho$, where conditions take precedence over assignments and calls.
\end{definition}

A patch-derived trigger line rests on the developer's fix and marks where the two versions start to diverge on $x$, which need not be where a fault would manifest, so the results that depend on it are reported separately. Of the \NumFunctions{} functions of the study, \NumObservedTrigger{} have a trigger line read from a fault report and \NumPatchDerived{} a patch-derived one, and Section~S14 of the supplementary material gives the detailed rules. The trigger line names one statement. The reason for the fault also involves the statements that produced the offending values and decided that the faulting statement executed, which dynamic slicing~\cite{korel1988dynamic,agrawal1990dynamic} captures for one run.

\begin{definition}[Dynamic dependence]
\label{def:dyndep}
Let the run $\rho$ be the sequence of executed instruction instances. An instance $j$ \emph{directly depends} on an earlier instance $i$ when one of the following holds: $j$ uses a register value that $i$ defined (data), $j$ accesses memory through an address that $i$ computed (address), $j$ reads a byte whose most recent write was $i$ (memory), the execution of $j$ was decided by the branch $i$, that is, the block of $j$ is control dependent on the block of $i$ in the post-dominator sense and $i$ is the most recent execution of such a branch (control), or $i$ and $j$ are a call and its callee's use of a parameter, or a return and the caller's use of the result (interprocedural).
\end{definition}

\begin{definition}[Dynamic slice]
Let $s_t$ be the instance of the faulting access on the trigger line, or, for a patch-derived trigger line, its last execution in $\rho$ (its first execution when $\rho$ does not terminate). The \emph{dynamic slice} $D\subseteq\mathit{lines}(\vv)$ is the set of lines of $\vv$ with an instruction instance from which $s_t$ is reachable through direct dependences, together with $t$.
\end{definition}

Dependence chains may leave $\vv$ through calls and return to it, and the slice keeps the lines of $\vv$ on such chains, because the detectors see only the function. Every line of $D$ was executed in $\rho$ and lies on a chain of dependences that ends in the faulting access, so agreement with $D$ cannot be reached by naming many lines.

\begin{ownproposition}
\label{prop:subset}
Let $S$ be a static backward slice of $\vv$ from the trigger line computed by a slicer that over-approximates the five kinds of dependence for all executions. Then $D\subseteq S$.
\end{ownproposition}

The proposition holds because every direct dynamic dependence is an occurrence in $\rho$ of a static dependence that such a slicer includes (Section~S14 of the supplementary material gives the proof). A line of the dynamic slice that is absent from a static slice therefore exposes a dependence the static slicer missed, for instance through an alias it does not model.

\parag{Scope.} The reference describes where a fault manifests and through which values and decisions the observed run reached it. A root cause that acts on the fault through allocation state, as in Example~\ref{ex:stale}, or that consists of code missing from the function~\cite{zhang2007omission}, lies outside it. Overlap with the reference is therefore a conservative measure of localization, and the localization test is complemented by tests that intervene on the program (Section~\ref{sec:design}).

\subsection{Construction and Selection}
\label{sec:obtaining}

We implement the reference on LLVM~\cite{lattner2004llvm}. The notion of dynamic dependence and the slicing algorithm are the established ones~\cite{agrawal1990dynamic,zhang2003precise} in the form implemented by Giri~\cite{sahoo2013giri}. Our pipeline adds the machinery that makes them applicable to sanitizer-detected faults of real programs at scale: a criterion derived from the fault report, tracing that coexists with sanitizers and survives the abort at the fault, a model of memory written by uninstrumented library code, and an explicit completeness verdict for every slice.

\parag{Proofs of concept.} BigVul and PrimeVul contain the text of each function and its fixing commit, and neither includes a triggering input or a harness. The harness $H$ and the input $x$ of every instance were generated by VulValidate~\cite{zhang2026vulvalidate}, in which Claude Opus 5.0 calls dynamic analysis tools on the revisions before and after the fix. The two revisions are rebuilt following ARVO's method~\cite{mei2026arvo}. The model only proposes the driver and the input, and their acceptance is mechanical. An instance counts as confirmed when the same input faults on the vulnerable build in two runs, the fault is attributed to the labeled function through its line range, and the fixed build runs clean, and our slicing pipeline replays every confirmed instance again in a fresh copy of the source tree before it enters the study. Each confirmed instance carries a self-contained evidence folder with the build script, the driver, both versions of the function, the input in its original bytes, and the logs of both runs.

\parag{Slicing and validation.} The pipeline replays the proof of concept in a private copy of the source tree and rebuilds both versions from the recovered commands. The criterion is the last executed instance of the access that the report names on the trigger line, so that a statement such as \texttt{dst[i] = src[i]} that reads out of bounds is sliced from its read. An LLVM pass placed before optimization and before the sanitizer instrumentation traces basic blocks, memory accesses, calls, and returns, and a backward search over the trace collects the instances on which the criterion depends. A slice is accepted as \emph{complete} only when the trace is untruncated and the run stopped at the criterion. In \HcdUbsan{} builds with UBSan, handler blocks without a successor add control dependences that Definition~\ref{def:dyndep} excludes, as in Example~\ref{ex:stale}. Recomputing control dependence without them changes \HcdChanged{} of the \HcdAnalysed{} slices that it processes within 120\,s each (\HcdNotAnalysed{} function exceeded this limit), removes \HcdLinesRemoved{} of their \HcdLines{} lines, and keeps the direction of every slice comparison (Section~S27 of the supplementary material), so the results use the slices as computed. We validated the slicer against 31 control programs with known dependences, each exercising one construct such as indirect calls, exceptions, or library callbacks, and against the original Giri implementation on LLVM~3.4 (Section~S10).

\parag{Selection.} Of the \NumSlicedInstances{} confirmed instances of BigVul~\cite{Fan2020} and PrimeVul~\cite{ding2024vulnerability} with a complete slice, \NumDistinctFunctions{} distinct functions remain after merging identical texts, and \NumCandidateFunctions{} of them have a trigger line and all static slices of Section~\ref{sec:static}. Before they are selected, two of the authors inspect the fixed version of every function, independently of each other and before any model output was seen, and a function becomes a candidate only when both judge its fixed version free of defects of any kind for the arguments its callers pass. Three dynamic checks of the fixed version follow. Its build must run clean on the trigger input, both in the evidence folder and in a rebuild of the developer's fix, which leaves out \NumExcludedFixedNotClean{} functions. The dynamic confirmation must name the labeled CVE, which leaves out \NumExcludedCveMatch{} functions whose confirmation triggered another real defect or states no judgment of the CVE. Further testing must reveal no defect or vulnerability attributable to the fixed function, which excludes \NumExcludedRecordCheck{} more. The tools follow the class of the fault, AddressSanitizer for \ToolASan{} functions, UndefinedBehaviorSanitizer for \ToolUBSan, LeakSanitizer for \ToolLSan, MemorySanitizer for \ToolMSan, ThreadSanitizer for \ToolTSan, assertions or timeouts for \ToolAssert{} and \ToolTimeout, and comparisons of outputs or authorization decisions for \ToolBehavior{} functions whose weakness causes no memory fault, and \ToolTwo{} functions use two tool families. Every fixed version runs twice on the trigger input in its evidence folder and twice in our rebuild, the runs show the fixed function executing for \FixedExecutedMain{} of the \NumFunctions{} functions, and \RcBeyondPocMain{} fixed versions also run on further attack inputs. The remaining \NumFunctions{} functions from \NumProjects{} projects, with CVE identifiers from \YearMin{} to \YearMax, form the study. \NumPairs{} of them have fixed text and form matched pairs, in which the fixed version is the developer's repair for the labeled CVE, labeled non-vulnerable by BigVul or PrimeVul, and we retain that label after the screening. The other \NumUnpaired, all from PrimeVul, have no fixed text in the dataset and enter only the measures that show the vulnerable version alone. Section~S10 of the supplementary material lists the tools and runs per function, and Table~S10 the full selection.

\subsection{Two Examples}
\label{sec:examples}

\begin{figure}[t]
\centering
\scriptsize
\setlength{\tabcolsep}{2.2pt}
\renewcommand{\arraystretch}{0.92}
\resizebox{\linewidth}{!}{%
\begin{tabular}{r@{\ \ }l@{\ }cccccc}
\multicolumn{8}{l}{\footnotesize\textbf{(a)} \texttt{seamless\_process}, rdesktop, CVE-2018-20182. Out-of-bounds read in the copy on line \ExampleReadTrigger.}\\[1pt]
\toprule
 & & \shortstack{Dynamic\\slice} & Joern & dg & \shortstack{Deep-\\Wukong} & Patch & \shortstack{GPT-6.1-sol\\evidence} \\
\midrule
\ExampleRead
\bottomrule
\\[-2pt]
\multicolumn{8}{l}{\footnotesize\textbf{(b)} \texttt{split}, shapelib, CVE-2022-0699. Double free on line \ExampleStaleTrigger.}\\[1pt]
\toprule
 & & \shortstack{Dynamic\\slice} & Joern & dg & \shortstack{Deep-\\Wukong} & Patch & \shortstack{GPT-6.1-sol\\evidence} \\
\midrule
\ExampleStaleShort
\bottomrule
\end{tabular}}
\caption{The functions of Examples~\ref{ex:read} and~\ref{ex:stale} with their references. $\blacktriangleright$ marks the trigger line. The columns Joern, dg, and DeepWukong are the static slices of these tools (DeepWukong backward to the trigger line). Each of the other columns repeats the number of every line that the reference contains. Patch: the number of every line that the fix modifies or deletes. The grey rows are the lines that the fix inserts, which exist only in the fixed version, and the Patch column gives their number there. The last column gives the lines that GPT-6.1-sol cites as evidence. In (b) lines 9 to 15 are elided, and Figure~S3 of the supplementary material gives the full listing. All measures use the full functions.}
\label{fig:examples}
\end{figure}

\begin{caseexample}[The fix guards the faulting access]
\label{ex:read}
In \texttt{seamless\_process} of rdesktop (Figure~\ref{fig:examples}a, CVE-2018-20182) the proof of concept passes a stream whose read pointer \texttt{s->p} already lies 8 bytes beyond \texttt{s->end}. Line 6 computes \texttt{pkglen = s->end - s->p}, which wraps around as an unsigned value, and the copy on line 9, the trigger line, reads beyond the receive buffer. The dynamic slice holds the length computation, the allocation sized by it, and the copy. The fix inserts a check of the stream state after line 4 (the grey lines in Figure~\ref{fig:examples}a), where the fixed build stops with a protocol error on the same input, so a reference built from fix lines would point before the computation of the faulting value. GPT-6.1-sol cites lines 3, 6, 8, 9, and 10, which contain the dynamic slice, and answers \ExampleReadAnsVuln{} for $\vv$ with confidence \ExampleReadConfVuln{} and \ExampleReadAnsFixed{} for $\ff$ with confidence \ExampleReadConfFixed.
\end{caseexample}

\begin{caseexample}[A limit of the reference]
\label{ex:stale}
In \texttt{split} of shapelib (Figure~\ref{fig:examples}b, CVE-2022-0699) the copy of the argument string is released on line 20. When the \texttt{realloc} on line 23 then fails, the error branch releases the copy again on line 29, which the fix deletes. The driver fails every \texttt{realloc} after the first, AddressSanitizer reports the double free at line 29 with stacks for both releases, and the fixed build runs clean. The dynamic slice follows the pointer \texttt{copy} from line 2 to line 29 and the decisions that lead there. Its lines 25 and 26 are an artifact of the implementation, since the UBSan check of \texttt{--i} on line 26 ends in a handler without a successor, which the post-dominator analysis takes for an exit, so that line 29 appears control dependent on the loop condition (Section~\ref{sec:obtaining}). The slice omits the first release on line 20, whose effect on line 29 passes through the state of the heap, which belongs to none of the five dependence kinds. Joern and DeepWukong contain line 20, and dg leaves it out. GPT-6.1-sol cites both releases and answers \ExampleStaleAnsVuln{} for the vulnerable version with confidence \ExampleStaleConfVuln. A reference that stops at the dynamic slice would count its citation of line 20 as an extra line, although that release is the root cause of the fault.
\end{caseexample}

\subsection{Comparison with Existing References}
\label{sec:static}

\parag{Fix lines.} On the \FixObsFunctions{} of the \CitesObsFunctions{} paired functions with an observed trigger line whose fix changes a line of $\vv$, the fix changes the trigger line itself in \FixObsTrigChanged\% of the functions, and in \FixObsAllAfter{} functions all changed lines lie after it. A reference built from patch lines alone can therefore miss a correctly identified fault site.

\parag{Static slices.} Static slices are the references of choice in prior work, and DeepWukong~\cite{Cheng:2021:DeepWuKong} uses them as model input. We compute three of them from the same criterion. The Joern slice~\cite{Joern} is the backward closure over data and control dependences of the program dependence graph of the labeled function, intraprocedural and without alias analysis. The dg slice~\cite{chalupa2020dg} is computed on LLVM bitcode with pointer analysis and dependences across functions. The DeepWukong slice is a corrected variant of its published procedure, which slices backward and forward from four kinds of program points and, for BigVul, also merges slices of the fixed function. Our variant slices backward only, from the program points up to the trigger line, in the vulnerable function (Section~S7 of the supplementary material).

\begin{table}[t]
\centering
\caption{Static slices against the dynamic slice. Retained: share of the dynamic slice that the static slice contains, mean with 95\% interval. Outside: share of the static slice outside the dynamic slice, mean. Size: size of the static slice over that of the dynamic slice, median. \LOName{}: $\LOsym(S,D)$ of Equation~\eqref{eq:jaccard}, median. Last column: share of functions whose static slice contains the whole dynamic slice (\%).}
\label{tab:staticdyn}
\small
\resizebox{\linewidth}{!}{%
\begin{tabular}{lrrrrr}
\toprule
Static slice & Retained (\%) & Outside (\%) & Size & \LOhead & \shortstack[r]{Contains the whole\\dynamic slice (\%)} \\
\midrule
\TabStaticDyn
\bottomrule
\end{tabular}}
\end{table}

Proposition~\ref{prop:subset} predicts a retention of 100\% for a sound slicer, and Table~\ref{tab:staticdyn} shows that every static slicer misses dependences that the fault exercised. dg retains the most, \SliceRetDg\% of the dynamic slice against \SliceRetJoern\% for Joern, a paired difference of \SliceDgMinusJoern{} percentage points, and it contains the whole dynamic slice for \SliceContainDg{} functions against \SliceContainJoern{} for Joern. The ordering dg, DeepWukong, Joern holds for each dataset separately and under stricter rules for matching the slicing criterion. In Figure~\ref{fig:examples}b, for instance, the Joern slice lacks the guard \texttt{if (i)} on line 22. Higher retention is bought with larger slices, since dg selects \SliceSizeDg\% of a function against \SliceSizeDyn\% for the dynamic slice, and \SliceExtraMin\% to \SliceExtraMax\% of the lines of every static slice lie outside the dependences of the observed run. The published DeepWukong procedure selects \SliceSizeDwkPub\% of a function and misses the trigger line in \SliceMissDwkPub\% of the functions, so it rewards any explanation that cites many lines, and restricting it to backward slices up to the trigger line raises its retention of the dynamic slice from \SliceRetDwkPub\% to \SliceRetDwk\% (Section~S9 of the supplementary material). The trigger line and the dynamic slice therefore serve as the primary references, and the static slices show which conclusions depend on the kind of reference.

\section{Study Design}
\label{sec:design}
\label{sec:ladder}

Table~\ref{tab:levels} lists the five levels. A model can exceed a baseline at one level and perform at chance at another, so each level is read against its own shortcut or control.

\subsection{Levels, Measures, and Baselines}

\parag{L1, discriminate.}\label{sec:lone} The detector sees one numbered function at a time and answers whether it contains a security vulnerability. Each matched pair consists of the confirmed vulnerable version $\vv$ and its screened fixed version $\ff$, and a vulnerable verdict on $\ff$ disagrees with the benchmark label. On balanced pairs a detector that always answers \emph{vulnerable} reaches an F1 of 0.67 without separating any pair (Table~S11 of the supplementary material lists further degenerate detectors).Our main measure is therefore Pair-wise Correct (P-C)~\cite{ding2024vulnerability}, the share of pairs in which $\vv$ is judged vulnerable and $\ff$ is judged not vulnerable. Accurate Differentiation Rate (ADR)~\cite{fse2026adr} uses the same criterion of requiring both judgments to be correct on matched SAT/UNSAT pairs. We report P-C alongside the share of pairs in which both versions are judged vulnerable (P-V). A detector that answers \emph{vulnerable} independently with probability $v$ reaches an expected P-C of $v(1-v)$, at most 0.25, the chance level of each model. Chance is a demanding baseline for a detector that gives both versions the same verdict, whose P-C stays far below it even when its other verdicts point the right way. We therefore also test the direction of the pairs on which the verdicts differ. On balanced pairs, recall minus the false-positive rate equals the share of correct minus that of reversed pairs, and an exact McNemar test with a Holm correction over the thirteen models compares the two counts. We also report the Matthews correlation coefficient (MCC)~\cite{chicco2020mcc}, which stays at 0 for every detector whose verdicts do not depend on the label. MCC and P-C answer different questions, since a detector can reach a positive MCC from functions that are easy across the board while separating few pairs. The AUC measures how well the model's score orders vulnerable above fixed versions, which is $\log P(\text{YES})-\log P(\text{NO})$ at the answer position for open-weight models read at the answer token and the stated confidence otherwise. The shortcut is a \emph{surface-feature classifier}, a logistic regression over ten features of one version, such as its length and its numbers of \texttt{if} statements, NULL checks, and risky calls, cross-fitted over five folds in which all functions of a CVE share a fold.

\parag{L2, localize.}\label{sec:ltwo} \emph{Stated evidence} is the list of line numbers a model returns with its verdict on $\vv$. \emph{Attributed evidence}, available for the open-weight models read at the answer token, ranks the lines by \LOOname{}, which blanks one line at a time and takes the drop of the model's score as the influence of that line~\cite{zeiler2014visualizing}, by gradient $\times$ input, or by integrated gradients~\cite{sundararajan2017axiomatic}. For a set $E$ of lines offered as evidence and a reference $T$, the trigger line or one of the slices, we report the coverage and the \LOname{},
\begin{equation}
\mathrm{Cov}(E,T) = \frac{|E\cap T|}{|T|}, \qquad \LOsym(E,T) = \frac{|E\cap T|}{|E\cup T|}, \label{eq:jaccard}
\end{equation}
per function and averaged. Coverage rises whenever more lines are cited, while the \LOname{} falls both when reference lines are omitted and when superfluous lines are added, so it is our main measure. For a ranking we take as many top-ranked lines as the reference has. Every value comes with its exact expectation for equally many random lines. Faults cluster around risky operations, so the stronger shortcut is the \emph{keyword rule}, which cites $|E|$ lines, those with a risky identifier such as \texttt{memcpy} or \texttt{malloc} first and random lines afterwards, and, for rankings, orders the lines by their number of risky identifiers.

\parag{L3, evidence sensitivity.}\label{sec:lthree} \emph{Evidence removal} blanks a set of lines of $\vv$, keeping line numbers and indentation, pairs the result with the unchanged $\ff$, and recomputes the P-C. The removed set is the dynamic slice, the DeepWukong slice, the trigger line alone, or as many random lines as the dynamic slice has, and one further condition keeps only the dynamic slice. \emph{Guidance} marks corresponding lines of both versions in the prompt, so the marking itself reveals nothing about which version is vulnerable, or appends the dynamic dependence edges as text. A set is compared with random lines of equal size where such a set exists and with the unchanged pair otherwise. Blanking changes content, syntax, and the presence of risky operations together, and an empty line where code is expected can itself act as a signal (Section~\ref{sec:rq4}). A third, observational check asks whether the pairs that a model separates are the pairs on which it cites the trigger line. L3 thus measures the sensitivity of the verdict to the reference evidence, which can differ from the lines a model states, and a test of whether the stated lines are the lines used would intervene on them with structurally matched controls.

\parag{L4, counterfactual.}\label{sec:lfour} The detector receives, one at a time and with the question of L1, variants of the function whose label is established by execution (Section~\ref{sec:benchmark}). Some edits keep the label, namely renaming all local variables and parameters on both versions and inserting a neutral statement into $\vv$. Others remove the fault, namely the developer's fix, an alternative fix written independently of it, and local fixes that apply one or a few hunks of the developer's patch. A \emph{decoy} adds a check of about the size of the developer's patch that looks like a fix and leaves the faulting statement unprotected, for example a check placed after that statement or on another variable. A decoy counts as faulting only when the proof of concept still faults on it in both runs. The comparisons are paired within functions, against the verdict on the real version or on the developer's fix, and a function passes the local-fix test when $\vv$ is judged vulnerable, a local fix not vulnerable, and the neutral edit vulnerable.

\parag{L5, execution.}\label{sec:lfive} The detector receives the driver sources that the build compiles, without comments and with conditional compilation resolved as built, the command line, the input bytes, and one version of the function with line numbers, either $\vv$, $\ff$, or a confirmed variant of L4. It answers whether this run reports a fault inside the function and on which line, and the truth is the confirmed outcome of that very run. A correct prediction is consistent with reasoning about this run, which is narrower than understanding the vulnerability in general. We report the accuracy of the outcome and the joint accuracy over faulting versions with a known faulting line, which requires a predicted fault and a stated line within two lines of it, the share of functions whose two real versions are both predicted correctly, and the localization among the versions predicted to fault, against the line with the most risky identifiers. A \emph{hidden-body} control sees the same driver, command line, and input with only the first line of the function. Two \emph{guard rules} test whether a visible bounds check explains the prediction. The local rule sees one version and predicts a fault when a risky line has no bounds check in the ten lines before it, and the added-guard rule knows $\vv$ and predicts a clean run exactly when the version adds a conditional statement. The driver and the input were written to trigger the fault, so a \emph{benign-input} control keeps the driver and uses an input on which $\vv$ runs clean.

\subsection{Models and Protocol}
\label{sec:setup}

\parag{Models.} The open-weight models are Qwen2.5-Coder-7B-Instruct and Qwen2.5-Coder-32B-Instruct~\cite{hui2024qwen25coder}, CodeLlama-34B-Instruct~\cite{roziere2023code}, the mixture-of-experts Qwen3-Coder-30B-A3B-Instruct~\cite{yang2025qwen3}, VulnLLM-R-7B~\cite{vulnllmr2025}, fine-tuned for vulnerability reasoning, Qwen3-Coder-Next with 8-bit weights, the dense Qwen3.6-27B~\cite{qwen36}, gpt-oss-20b~\cite{gptoss2025}, and Foundation-Sec-8B-Reasoning~\cite{foundationsec2025reasoning}, fine-tuned for security reasoning, all at pinned revisions. The five models without a reasoning mode are read at the answer token. The four reasoning models, VulnLLM-R-7B, Qwen3.6-27B with its thinking mode on, gpt-oss-20b with high reasoning effort, and Foundation-Sec-8B-Reasoning, reason first and answer in the JSON format of the commercial models, with a reasoning budget of up to \ReasonBudgetGptOss{} tokens (Section~S25 of the supplementary material). Attribution, prompt paraphrases, and the linear head need the score at the answer position and use runs read at the answer token, available for VulnLLM-R-7B and Qwen3.6-27B. The commercial models GPT-6.1-sol and GPT-5.6-sol are queried through the Codex command-line client, and Claude Opus 5.5 and Claude Opus 4.8 through the Claude Code command-line client, all four with high reasoning effort. Each query is a single question whose answer is constrained by a JSON schema, and no model can use a tool, search the web, or read a local file. The GPT clients run with a restricted filesystem and with tools and connectors disabled, and none of the \IsoGptQueries{} GPT transcripts contains a tool call. Every commercial reply is checked for the model name and reasoning effort that served it. L1 and L2 cover all thirteen models, guidance eleven of them, and evidence removal, L4, and L5 the four commercial models. Section~S11 of the supplementary material gives revisions, client versions, and the number of functions per experiment and model.

\parag{Prompts.} The detection prompt shows one function with line numbers and asks for YES or NO. Commercial and reasoning models additionally state a confidence. The explanation prompt asks for the verdict, the line numbers the decision is based on, and a short justification.

\subsection{CVEs Published After the Training Cutoffs}
\label{sec:postcut}

The CVEs of the pre-cutoff set date from \YearMin{} to \YearMax, before every training cutoff, so exposure to the code during training~\cite{carlini2021extracting,Yang2024} is a possible explanation for any result. We therefore repeat the experiments on vulnerabilities that became public only after the commercial models were trained. The vendors state a training cutoff of June 2026 for Claude Opus 5.5~\cite{anthropic2026opus55}, 30 April 2026 for GPT-6.1-sol~\cite{openai2026gpt61sol}, 16 February 2026 for GPT-5.6-sol~\cite{openai2026gpt56sol}, and January 2026 for Claude Opus 4.8~\cite{anthropic2026opus48}. From the CVE list of cve.org we sampled CVEs published on or after 1 September 2026, two months after the latest of these cutoffs, in open-source C and C++ projects whose public fix changes a C or C++ function. The confirmation method of the pre-cutoff set reproduced a fault in the labeled function for \PostCutConfirmed{} of them, and \PostCutStudyCVEs{} pass the screening of Section~\ref{sec:obtaining}. They come from \PostCutProjects{} projects, were published between \PostCutDateFirst{} and \PostCutDateLast{} with identifiers from \PostCutCVELow{} to \PostCutCVEHigh, and all their fix commits were committed between \PostCutFixFirst{} and \PostCutFixLast, after every cutoff. L2 and L3 use the \PostCutFunctions{} of them with all slices, and L1 is reported on both sets (Table~S10 of the supplementary material).

\subsection{Statistical Procedure}
\label{sec:stats}

Functions that share a CVE are dependent. Unless stated otherwise, intervals are 95\% percentile intervals from 2000 bootstrap draws that resample CVEs~\cite{efron1993bootstrap}. When every value of a cell is the same, the bootstrap interval collapses to a point, and the cell gives the 95\% Wilson score interval computed on the number of CVEs behind the cell. A difference is significant when its interval excludes zero. Comparisons across models use the functions that all compared models answered, and paired differences are formed within functions before resampling. In the tables, $\uparrow$ marks a difference whose interval lies entirely above zero, $\downarrow$ one whose interval lies entirely below zero, and $\approx$ one whose interval contains zero.

\section{RQ1: Discriminating Vulnerable from Fixed Functions}
\label{sec:rq1}

\begin{table}[t]
\centering
\caption{L1, pair discrimination on up to \NumCommon{} matched pairs, open-weight models grouped by protocol. Chance is the independent-guessing P-C at each model's share of \emph{vulnerable} verdicts. Differences in percentage points with 95\% intervals. $^{\dagger}$CVE-grouped cross-fitted baseline.}
\label{tab:pairs}
\small
\setlength{\tabcolsep}{4.5pt}
\resizebox{\linewidth}{!}{%
\begin{tabular}{lrrrrrrrrr}
\toprule
Model & Acc. & Prec. & Recall & F1 & MCC & AUC & P-C & Chance & P-C $-$ chance \\
\midrule
\TabPairs
\bottomrule
\end{tabular}}
\end{table}

\parag{Open-weight models.} Table~\ref{tab:pairs} reports the pair discrimination. The surface-feature classifier, which sees only length and counts of syntactic constructs, reaches a P-C of \SurfaceADR\% and an MCC of \SurfaceMCC. Of the five open-weight models without a reasoning mode, only Qwen2.5-Coder-32B exceeds this classifier, and none exceeds its chance level. A high F1 can hide this, since CodeLlama-34B labels almost every version vulnerable and combines an F1 of \CodeLlamaFOne{} with a P-C of \CodeLlamaADR\%. The four reasoning models exceed the surface classifier, yet none of them exceeds its chance level either, with differences of \ChanceGapVulnLLM, \ChanceGapQwenThreeSix, \ChanceGapGptOss, and \ChanceGapFoundationSec{} points, because they often give both versions the same verdict. The scores of the two versions of a function differ far less than those of different functions, so a single threshold gives both versions the same label.

\parag{Direction of the separated pairs.} Chance, however, hides a signal in these models. Among the pairs on which a model gives the two versions different verdicts, \DirSigOpen{} of the \DirOpenModels{} open-weight models judge the vulnerable version vulnerable significantly more often than the fixed one (exact McNemar test, Holm-corrected). Correct minus reversed pairs amounts to \DirCRQwenThreeSix{} points for Qwen3.6-27B, \DirCRGptOss{} for gpt-oss-20b, \DirCRVulnLLM{} for VulnLLM-R-7B, and \DirCRQwenThirtyTwo{} and \DirCRQwenThree{} for Qwen2.5-Coder-32B and Qwen3-Coder-30B, while Qwen3-Coder-Next, Foundation-Sec-8B-Reasoning, Qwen2.5-Coder-7B, and CodeLlama-34B show no significant direction after the correction. These models thus carry a weak sensitivity to the fix, and their habit of giving both versions the same verdict keeps their P-C at the chance level (Figure~\ref{fig:pairs}).

\parag{Commercial models.} The commercial models exceed the surface classifier by \SurfaceGapSixOneCI{} points (GPT-6.1-sol), \SurfaceGapFiveSixCI{} (GPT-5.6-sol), \SurfaceGapClaudeCI{} (Claude Opus 5.5), and \SurfaceGapClaudeFourEightCI{} (Claude Opus 4.8), and three of them exceed their chance level, all but GPT-5.6-sol. For all four, far more of the pairs with differing verdicts are correct than reversed, with correct minus reversed pairs between \DirCRFiveSix{} points (GPT-5.6-sol) and \DirCRClaude{} (Claude Opus 5.5). The GPT models judge both versions vulnerable in \GptSixOneBothVuln\% and \GptFiveSixBothVuln\% of the pairs, and their P-C falls with the length of the function (Section~\ref{sec:rq4}). Claude Opus 5.5 separates \ClaudeADR\% of the pairs, a lead that Section~\ref{sec:rq4} examines against recognition of known vulnerabilities.

\begin{figure}[t]
\centering
\FigPairOutcomes
\caption{L1 outcome of the matched pairs that all models answered, in the four categories of PrimeVul~\cite{ding2024vulnerability}. A pair is correct (P-C) when the vulnerable version is judged vulnerable and the fixed version not vulnerable, and reversed (P-R) in the opposite case. Models are grouped as in Table~\ref{tab:pairs}.}
\label{fig:pairs}
\end{figure}

\parag{Where the pairs go.} Figure~\ref{fig:pairs} shows how the pairs that a model fails to separate are distributed. The commercial models rarely reverse a pair, so their losses come from giving both versions the same verdict, almost always \emph{vulnerable}. For the GPT models these pairs outnumber the correct ones, while Claude Opus 5.5 judges both versions vulnerable for \ClaudeBothVuln\% of the pairs. The open-weight models fail in opposite directions. Qwen2.5-Coder-7B judges both versions not vulnerable for most pairs, CodeLlama-34B and Foundation-Sec-8B-Reasoning judge both vulnerable, and the reasoning models reverse more pairs than any commercial model. A detector can therefore lose a pair in two opposite ways, and only the pair-level view shows which.

\parag{Prompts and hidden states.} Under two paraphrased templates (Table~S1 of the supplementary material) the P-C of Qwen2.5-Coder-32B is stable in aggregate, yet only \TplAllThree\% of the pairs are separated under all three templates while \TplAny\% are separated under at least one, the signature of noise around a weak signal. A linear head on the hidden state at the answer position, cross-fitted over functions, raises the AUC of Qwen2.5-Coder-32B only from \HeadThirtyTwoAUCbefore{} to \HeadThirtyTwoAUCafter{} (Table~S2). The hidden state thus carries a weak ranking signal and no usable decision.

\parag{Answer to RQ1.} No model reliably tells a vulnerable function from its screened fix. The open-weight models stay at chance despite a weak sensitivity to the fix, and the commercial models, three of which exceed chance, still label many fixes vulnerable.

\section{RQ2: Does the Evidence Point to the Fault?}
\label{sec:rq2}

\begin{table}[t]
\centering
\caption{L2, how often the cited lines contain the trigger line (\% of functions), for the model and for two rules that pick equally many lines per function, the keyword rule (lines with a risky identifier such as \texttt{memcpy} first, then random lines) and random lines. Each model on all functions it answered. Last two columns: model minus keyword rule in percentage points with 95\% interval, and the arrow of Section~\ref{sec:stats}.}
\label{tab:keyword}
\small
\begin{tabular}{lrrrrc}
\toprule
Model & \shortstack[r]{Cited by\\the model} & \shortstack[r]{Keyword\\rule} & Random & \shortstack[r]{Model minus\\keyword rule} & \\
\midrule
\TabKeyword
\bottomrule
\end{tabular}
\end{table}

\parag{Stated evidence.} Table~\ref{tab:keyword} reports how often the cited lines contain the trigger line. The trigger line is often a visibly risky statement, since \KeywordHasTrigger\% of the trigger lines contain one of the keywords, and the keyword rule covers it about twice as often as random lines. The four commercial models cite four to five lines and exceed the keyword rule by \KwGapClosedMin{} to \KwGapClosedMax{} percentage points, so their localization goes beyond the selection of risky-looking statements. Restricted to the \CitesObsFunctions{} functions whose trigger line is observed in a fault report, the margins are \SixOneKeywordGapObs, \FiveSixKeywordGapObs, \ClaudeKeywordGapObs, and \ClaudeFourEightKeywordGapObs{} points for GPT-6.1-sol, GPT-5.6-sol, Claude Opus 5.5, and Claude Opus 4.8, and on patch-derived trigger lines they remain positive (Table~S6 of the supplementary material). Among the open-weight models, Qwen3.6-27B and gpt-oss-20b exceed the rule by \KeywordGapCIQwenThreeSix{} and \KeywordGapCIGptOss{} points, the two largest margins, while VulnLLM-R-7B, Foundation-Sec-8B-Reasoning, Qwen2.5-Coder-7B, and CodeLlama-34B are indistinguishable from it. The weaker open-weight models fail in two ways. CodeLlama-34B and VulnLLM-R-7B cite about nine lines per function, which raises their chance of containing the trigger line, and the two Qwen2.5-Coder models and Qwen3-Coder-30B cite about one line and often return an empty list.

\parag{Agreement with the slices.} On the \NumSelfCommon{} functions that nine models explained (Table~\ref{tab:statedmain}), the \LOname{} of the commercial models with every reference exceeds its random expectation by a factor of two to five. Paired within functions, their cited lines match the dynamic slice better than the Joern and DeepWukong slices and about as well as the dg slice, the static slice that retains most of the dynamic slice. For GPT-6.1-sol the \LOname{} with the dynamic slice exceeds that with Joern by \JaccDminusJoernSixOne{} and that with DeepWukong by \JaccDminusDwkSixOne, while the difference to dg is \JaccDminusDgSixOne, and it reaches \SixOneJaccD{} with the dynamic slice and \SixOneJaccW{} with the DeepWukong slice. The absolute values remain moderate (at most \ClaudeJaccD), because the models cite about four lines while a dynamic slice holds more.

\begin{table}[t]
\centering
\caption{L2, stated evidence against every reference, on the \NumSelfCommon{} functions that the nine models other than Qwen3-Coder-Next, Qwen3.6-27B, gpt-oss-20b, and Foundation-Sec-8B-Reasoning explained. Trigger line cited: share of functions whose cited lines contain the trigger line (\%). \LOName{}: $\LOsym(E,T)$ of Equation~\eqref{eq:jaccard}, with the expectation for equally many random lines in parentheses. Bold: significantly above that expectation. Union: the lines in the dynamic slice or in the static slice of Joern or dg.}
\label{tab:statedmain}
\small
\setlength{\tabcolsep}{3pt}
\resizebox{\linewidth}{!}{%
\begin{tabular}{lccccccc}
\toprule
& & \multicolumn{6}{c}{\LOName{} with} \\
\cmidrule(lr){3-8}
Model & \shortstack{Trigger line\\cited} & \shortstack{Trigger\\line} & \shortstack{Dynamic\\slice} & \shortstack{Joern\\slice} & \shortstack{dg\\slice} & \shortstack{DeepWukong\\slice} & Union \\
\midrule
\TabStatedJaccard
\bottomrule
\end{tabular}}
\end{table}

Table~\ref{tab:statedmain} also shows how the reference shapes the conclusion. Every model reaches a lower \LOname{} with the DeepWukong slice than with the dynamic slice, because that slice contains many lines outside the dependences of the fault. The models that cite many lines, CodeLlama-34B and VulnLLM-R-7B, gain most against the large dg slice and the union, while the commercial models reach about the same value with the dg slice as with the dynamic slice. A large static reference thus rewards citing many lines, and the dynamic slice removes this reward.

\parag{Attributed evidence.} The attribution methods of the open-weight models rank the trigger line first for \RankedTopOneMin\% to \RankedTopOneMax\% of the functions, against 4\% by chance, while the ranking by risky keywords does so for \KwTopOneMin\% to \KwTopOneMax\%, and \AttrBelowKwTopOne{} of the \AttrCombos{} attribution rankings fall significantly below it. The best ranking, \LOOname{} of VulnLLM-R-7B, reaches the keyword ranking in \LOname{} with the dynamic slice (\VulnOcclJaccD{} against \VulnOcclKwJaccD), and \LOOname{} of the Qwen models overlaps with the dynamic slice significantly \emph{below} chance, so the lines whose removal moves their score most tend to lie outside the dependences of the fault. Against the slices at a budget equal to their size, \RankedUp{} of \RankedCells{} combinations of model, method, and slice lie significantly above the random expectation and \RankedDown{} below it, and the attributions miss \RankedMissMin\% to \RankedMissMax\% of the dynamic slice. Larger budgets leave this picture unchanged (Tables~S3 and~S13 and Figure~S1 of the supplementary material).

\parag{Answer to RQ2.} The commercial models and two open-weight reasoning models locate the fault beyond the keyword shortcut, while the internal attributions of the open-weight models reach at most its level.

\section{RQ3: Do Verdicts Depend on the Evidence?}
\label{sec:rq3}

If correct verdicts stem from correct reasons, the functions on which a model separates the pair should be the functions on which it localizes the fault. For GPT-6.1-sol the share of functions with a cited trigger line is \NestCorrect{} percent among the correctly separated pairs and \NestWrong{} percent among the others, and the same ordering holds for every commercial model, with overlapping intervals only for Claude Opus 5.5. The interventions test this coupling directly.

\begin{table}[t]
\centering
\caption{L3, evidence removal for the four commercial models. Lines of the vulnerable version are blanked and the result is paired with the unchanged fixed version. P-C in percent, and the P-C difference in percentage points with a 95\% interval, for the dynamic slice to blanking as many random lines and for the other conditions to the unchanged pair. Table~S16 of the supplementary material adds the confidence ordering.}
\label{tab:blank}
\footnotesize
\setlength{\tabcolsep}{3.5pt}
\begin{tabular}{lrlrl}
\toprule
\multicolumn{5}{l}{\emph{(a) GPT models}} \\
\TabBlankHeadGPT
\midrule
\TabBlankGPT
\midrule
\multicolumn{5}{l}{\emph{(b) Claude models}} \\
\TabBlankHeadClaude
\midrule
\TabBlankClaude
\bottomrule
\end{tabular}
\end{table}

\parag{Evidence removal.} Relative to blanking equally many random lines, blanking the dynamic slice changes the P-C of GPT-6.1-sol by \MaskDvsRandomSixOne{} percentage points, that of GPT-5.6-sol by \MaskDvsRandomFiveSix, and that of Claude Opus 4.8 by \MaskDvsRandomClaudeFourEight{} (Table~\ref{tab:blank}). The random comparison matters, because blanking by itself shifts the verdicts. Random blanking raises the P-C of GPT-5.6-sol by \MaskRandomVsNoneFiveSix{} points relative to the unchanged function, and that of Claude Opus 4.8 from \MaskADRNoneClaudeFourEight\% to \MaskADRRandomClaudeFourEight\%. Claude Opus 5.5 behaves differently. With the dynamic slice blanked it still labels \MaskYesBlankDClaude\% of the vulnerable versions vulnerable, against \MaskYesNoneClaude\% unchanged, and its P-C changes by \MaskDvsRandomClaude{} points. Its confidence responds more than its verdict. The vulnerable version receives the higher confidence in \MaskOrderedNoneClaude\% of the unchanged pairs, \MaskOrderedRandomClaude\% after random blanking, and \MaskOrderedBlankDClaude\% after blanking the dynamic slice. The verdict persists also on functions whose CVE it does not name (\MaskYesBlankDUnnamedClaude\% against \MaskYesBlankDNamedClaude\% for named CVEs). On the CVEs published after its training, the same removal lowers its P-C (Section~\ref{sec:rq4}).

\begin{table}[t]
\centering
\caption{L3, marking lines in the prompt. P-C (\%) without marks, with the trigger line marked, and with random lines marked, as many as the union of the slices has. Then P-C differences in percentage points with 95\% intervals, for the trigger line and the dynamic slice against no marks and for the union of the slices against random marking of equal size. Table~S4 of the supplementary material gives all eight conditions.}
\label{tab:markmain}
\small
\setlength{\tabcolsep}{3.5pt}
\resizebox{\linewidth}{!}{%
\begin{tabular}{lrrrlll}
\toprule
Model & No marks & \shortstack[r]{Trigger line\\marked} & \shortstack[r]{Random lines\\marked} & \shortstack{Trigger line\\minus no marks} & \shortstack{Dynamic slice\\minus no marks} & \shortstack{Union\\minus random} \\
\midrule
\TabMark
\bottomrule
\end{tabular}}
\end{table}

\parag{Guidance.} Table~\ref{tab:markmain} reports the guidance conditions. Marking the union of the slices raises the P-C over marking as many random lines by \MarkSvsRandomSixOne{} percentage points for GPT-6.1-sol, and the gains of the other three commercial models include zero. Marking the trigger line raises the P-C of GPT-6.1-sol by \MarkVvsPlainSixOne{} and that of GPT-5.6-sol by \MarkVvsPlainFiveSix{} points relative to the unmarked pair, while marking random lines changes it by \MarkRandomvsPlainSixOne{} and \MarkRandomvsPlainFiveSix{} points, so the gain comes from the line and not from the act of marking. Marking the dynamic slice changes the P-C of every commercial model with an interval that includes zero. Appending the dependence edges as text helps GPT-6.1-sol (\EdgesVsShuffledSixOne{} points over shuffled edges). Among the open-weight models, marking the trigger line or the dynamic slice helps significantly only Qwen2.5-Coder-7B, from a P-C near zero (Tables~S4 and~S14 of the supplementary material).

\parag{Answer to RQ3.} The verdicts depend only in part on the evidence. Removing the dynamic slice lowers the pair discrimination of three commercial models and leaves that of Claude Opus 5.5 unchanged before the cutoffs, and supplying the evidence helps little.

\section{RQ4: Verdicts on Execution-Confirmed Edits}
\label{sec:benchmark}

L4 and L5 need variants of each function whose labels are established by execution. A variant replaces the function in a private copy of the source tree, the program is rebuilt with the recovered build commands, and the proof of concept runs twice. The variant is labeled faulting when the oracle fires in both runs and clean when both runs complete without any report, and other outcomes are discarded. GPT-6.1-sol writes the alternative fixes in both sets of functions from the vulnerable function, the developer's patch, and the kind and line of the observed fault, and it writes the decoys of the pre-cutoff set, while Claude Opus 5.5 writes those of the post-cutoff CVEs (Section~S22 of the supplementary material).

\parag{Benchmark.} Candidates require fixed text, a build in which the function is in its own file, a line-level oracle, and at most 150 lines. Of \CohortSize{} candidates selected with priority for distinct CVEs, \CtwoSixOneFunctions{} functions pass the baseline checks and the selection of Section~\ref{sec:obtaining}. On them execution confirmed \VariantsConfirmed{} variants, among them \AltFixFunctions{} alternative fixes, \DecoyFaultFunctions{} decoys that still fault, and \DecoyCleanFunctions{} decoys that removed the fault, and both renamed versions are confirmed for \RenameConfirmedFunctions{} functions. GPT-6.1-sol thus evaluates edits it generated, whereas the other models had no part in generating them.

\begin{table}[t]
\centering
\caption{L4, verdicts on execution-confirmed variants of each function. Each row names the version shown to the model and the verdict that is correct for it, and each cell gives the percentage of functions for which the model gives that verdict, with a 95\% interval. In (b) a verdict is correct when it equals the verdict on the unrenamed version. The row marked $\Delta$ gives the difference between the two rows above it on the same functions, in percentage points. Every function of this benchmark comes from a different CVE, so equal shares show identical intervals.}
\label{tab:lfour}
\small
\setlength{\tabcolsep}{3pt}
\resizebox{\linewidth}{!}{%
\TabLfourBegin
\toprule
\TabLfourHead \\
\midrule
\multicolumn{6}{l}{\emph{(a) Real versions, each shown alone, all functions}} \\
Vulnerable version & vulnerable & 94.5 [89.9, 98.2] & 81.7 [74.3, 88.1] & 97.2 [93.6, 100.0] & 76.1 [67.9, 83.5] \\
Fixed version & not vulnerable & 46.8 [37.6, 56.9] & 40.4 [31.2, 50.5] & 80.7 [73.4, 88.1] & 67.0 [57.8, 75.3] \\
Both versions (P-C) & both right & 42.2 [33.0, 51.4] & 25.7 [17.4, 33.9] & 78.9 [71.6, 86.2] & 44.0 [34.9, 53.2] \\
\midrule
\multicolumn{6}{l}{\emph{(b) Renamed versions, functions renamed in both versions}} \\
Renamed vulnerable version & as on the original & 95.8 [91.7, 99.0] & 96.9 [92.7, 100.0] & 96.9 [92.7, 100.0] & 87.5 [80.2, 93.8] \\
Renamed fixed version & as on the original & 89.6 [83.3, 94.8] & 89.6 [83.3, 95.8] & 89.6 [83.3, 94.8] & 83.3 [76.0, 90.6] \\
Original pair (P-C) & both right & 41.7 [32.3, 51.0] & 25.0 [16.7, 33.3] & 82.3 [74.0, 89.6] & 45.8 [35.4, 55.2] \\
Renamed pair (P-C) & both right & 38.5 [29.2, 47.9] & 26.0 [17.7, 35.4] & 81.2 [72.9, 88.5] & 46.9 [36.5, 56.2] \\
\midrule
\multicolumn{6}{l}{\emph{(c) Fixes, each shown alone, functions whose vulnerable version is judged vulnerable}} \\
Developer's fix & not vulnerable & 42.7 [32.6, 52.8] & 32.5 [22.1, 42.9] & 83.3 [75.6, 91.1] & 58.1 [45.9, 68.9] \\
Alternative fix & not vulnerable & 52.8 [41.6, 62.9] & 45.5 [33.8, 57.1] & 81.1 [72.2, 88.9] & 63.5 [51.4, 74.3] \\
$\Delta$ alternative minus developer's fix & & 10.1 [$-1.1$, 21.3] & 13.0 [3.9, 23.4] & $-2.2$ [$-11.1$, 6.7] & 5.4 [$-5.4$, 16.2] \\
\bottomrule
\end{tabular}}
\end{table}

\parag{Renaming and fixes.} Renaming hardly matters (Table~\ref{tab:lfour}b). The four models keep their verdict on the renamed vulnerable version for \CtwoSixOneRenameSame\%, \CtwoFiveSixRenameSame\%, \CtwoClaudeRenameSame\%, and \CtwoClaudeFourEightRenameSame\% of the functions, so the verdicts survive the loss of identifier names and of the verbatim text, and the P-C on the renamed pairs stays within a few points of that on the original pairs. The fixes separate the models (Table~\ref{tab:lfour}c). Among the functions whose vulnerable version they judge vulnerable, the GPT models accept only \CtwoSixOneDevFixNo\% and \CtwoFiveSixDevFixNo\% of the developer's fixes and accept the alternative fixes somewhat more often, a difference that is significant only for GPT-5.6-sol, which took no part in writing them. Claude Opus 5.5 accepts both kinds of fix alike, \CtwoClaudeDevFixNo\% and \CtwoClaudeAltFixNo\%, and Claude Opus 4.8 lies between them, with \CtwoClaudeFourEightDevFixNo\% and \CtwoClaudeFourEightAltFixNo\%.

\parag{Local fixes.} The fix-intervention stage of our slicing tool splits the developer's patch into hunks and builds $\vv$ with one hunk, or the smallest set of two or three hunks, applied. Such a variant becomes a \emph{local fix} when the trigger input faults on $\vv$ and no longer on the variant in two runs each, gdb shows the changed lines executing under the trigger, a control input that executes the function behaves as on $\vv$, and a neutral edit that inserts \texttt{volatile int vv\_neutral = 0;} into $\vv$ still faults. Local fixes cover \DpsCommon{} functions of the study, chosen independently of those of L4 and L5. Among the functions whose vulnerable version is judged vulnerable, GPT-6.1-sol accepts the local fix for \DpsSixOneFlips\%, GPT-5.6-sol for \DpsFiveSixFlips\%, Claude Opus 5.5 for \DpsClaudeFlips\%, and Claude Opus 4.8 for \DpsClaudeFourEightFlips\% (Table~S15 of the supplementary material), and neutral edits leave the verdict unchanged for more than 90\% of the functions in every model except Claude Opus 4.8 (\DpsClaudeFourEightNeutral\%). Local fixes are confirmed on the trigger input and on a control input, which falls short of a guarantee of safety for every input, so a vulnerable verdict on them can be justified for other inputs. Qwen2.5-Coder-7B, the one open-weight model evaluated on these variants, judges only \DpsQwenSevenVuln\% of the vulnerable versions vulnerable and rarely has a verdict to revise.

\parag{Answer to RQ4.} Verdicts are robust to renaming and neutral edits, while the acceptance of execution-confirmed repairs varies widely across models.

\section{RQ5: Predicting One Execution}
\label{sec:rq5}

\newcommand{\cfcode}[2]{\makebox[1.2em][r]{\tiny #1}\ \ \texttt{#2}}
\newcommand{\cfw}[1]{\textbf{#1}}
\begin{figure}[t]
\centering
\scriptsize
\setlength{\tabcolsep}{3pt}
\renewcommand{\arraystretch}{0.95}
\begin{minipage}{\linewidth}
{\footnotesize \texttt{findNextBorderPixel}, leptonica, CVE-2020-36278. Heap buffer overflow (read) on line 20. The input describes an 8$\times$8 image and starts the search at its corner pixel (0,\,0).}\\[2pt]
\cfcode{1}{findNextBorderPixel(l\_int32 w, l\_int32 h, l\_uint32 *data, l\_int32 wpl,}\\
\cfcode{}{\ \ \ \ \textrm{\textit{lines 2 to 13 take the start pixel (px, py) and three output pointers and declare the locals}}}\\
\cfcode{14}{\ \ \ \ qpos = *pqpos;}\\
\cfcode{15}{\ \ \ \ for (i = 1; i < 8; i++) \{}\\
\cfcode{16}{\ \ \ \ \ \ \ \ pos = (qpos + i) \% 8;}\\
\cfcode{17}{\ \ \ \ \ \ \ \ npx = px + xpostab[pos];}\\
\cfcode{18}{\ \ \ \ \ \ \ \ npy = py + ypostab[pos];}\\
\cfcode{}{\textcolor{gray}{\ \ \ \ \ \ \ \ if (npx < 0 || npx >= w || npy < 0 || npy >= h)\ \ \ \textrm{\textit{line 19 of the fixed version}}}}\\
\cfcode{}{\textcolor{gray}{\ \ \ \ \ \ \ \ \ \ \ \ continue;\ \ \ \textrm{\textit{line 20 of the fixed version}}}}\\
\cfcode{19}{\ \ \ \ \ \ \ \ line = data + npy * wpl;}\\
\cfcode{20}{\ \ \ \ \ \ \ \ val = GET\_DATA\_BIT(line, npx);}\\
\cfcode{}{\ \ \ \ \ \ \ \ \textrm{\textit{lines 21 to 30 return the neighbor when its bit is set, and 1 when none is}}}\\[3pt]
{\small
\setlength{\tabcolsep}{2pt}
\renewcommand{\arraystretch}{1.06}
\begin{tabular*}{\linewidth}{@{\extracolsep{\fill}}llcccccc@{}}
\toprule
 & & & & \multicolumn{4}{c}{L4 verdict (function only)} \\
\cmidrule(lr){5-8}
Version & Change & Run & L5
  & \shortstack{GPT-\\6.1-sol} & \shortstack{GPT-\\5.6-sol}
  & \shortstack{Claude\\Opus 5.5} & \shortstack{Claude\\Opus 4.8} \\
\midrule
Vulnerable & Original & Fault (20) & Fault (20) & V & V & V & V \\
Developer's fix & Guard before line~19 & Clean & Clean & N & N & N & N \\
Alternative fix & Read only in bounds & Clean & Clean & N & N & N & N \\
Decoy & Guard after line~20 & Fault (20) & Fault (20) & V & V & V & V \\
Renamed vuln. & Rename identifiers & Fault (20) & Fault (20) & V & V & V & V \\
Renamed fix & Rename identifiers & Clean & Clean & N & N & N & N \\
\bottomrule
\end{tabular*}}
\end{minipage}
\caption{Six counterfactual variants of one function. Run gives the measured outcome, and L5 gives the unanimous prediction with the driver and input. Parentheses give the faulting line. The grey lines are the guard that the developer's fix inserts, and they carry their numbers in the fixed version. Every other line number refers to the vulnerable version. At L4, V and N mean vulnerable and not vulnerable.}
\label{fig:cf}
\end{figure}

\begin{caseexample}[One function at L4 and at L5]
\label{ex:lfive}
In Figure~\ref{fig:cf}, leptonica's \texttt{findNextBorderPixel} (CVE-2020-36278) searches eight neighbors computed from two offset tables. The driver supplies the image size, starting pixel, and search direction. With the corner input, the first neighbor lies outside the image, and AddressSanitizer reports a heap buffer overflow at the read on line~20. The developer's fix checks both coordinates before the read (the grey lines in Figure~\ref{fig:cf}) and passes the additional dynamic screening. The alternative fix reads only in bounds. The decoy applies the developer's check after the read, so the fault persists. All four models judge all six versions correctly at L4 and predict every outcome and faulting line at L5.
\end{caseexample}

\begin{table}[t]
\centering
\caption{L5, prediction of one execution. The overall rows pool all evaluated versions with confirmed outcomes. Brackets give 95\% intervals where estimated. The hidden-body control sees the function's first line with the same driver and input. The benign-input control keeps the driver and uses an input on which $\vv$ runs clean. \emph{Both real versions} requires both predictions for a pair to be correct. The joint measure requires a fault prediction and a line within two lines of the observed fault site. The rows by kind of version show where the errors fall.}
\label{tab:cthreea}
\small
\setlength{\tabcolsep}{3pt}
\resizebox{\linewidth}{!}{%
\TabCthreeABegin
\toprule
Measure & \TabCthreeAHead \\
\midrule
\TabCthreeA
\bottomrule
\end{tabular}}
\end{table}

\parag{Outcome prediction.} Table~\ref{tab:cthreea} reports the predictions on \CthreeASixOneQuestions{} versions of \CthreeASixOneFunctions{} functions. The four models predict the outcome correctly for \CthreeAAccMin\% to \CthreeAAccMax\% of the versions, with MCC values of \CthreeASixOneMCC, \CthreeAFiveSixMCC, \CthreeAClaudeMCC, and \CthreeAClaudeFourEightMCC, while answering \emph{fault} for every version scores \CthreeAAlwaysFault\%. Example~\ref{ex:lfive} shows one function on which they predict every version, including the decoy whose check sits after the faulting read. Their errors fall on the vulnerable version, which they predict to fault in \CthreeAVzeroClosedMin\% to \CthreeAVzeroClosedMax\% of the functions, while the fixed versions and the alternative fixes are almost always predicted clean. Over the faulting versions with a known faulting line, the joint fault-and-location accuracy lies between \CthreeAJointClosedMin\% and \CthreeAJointClosedMax\%, and when GPT-6.1-sol predicts a fault, it names a line within two lines of the faulting line for \CthreeASixOneLine\% of the versions, against \CthreeASixOneKeywordLine\% for the line with the most risky tokens.

\parag{Against the guard rules.} The local guard rule reaches \GuardLocalAcc\%, because it predicts a fault for \GuardLocalFaultShare\% of the versions. The added-guard rule, which knows $\vv$ and is the strongest form of the explanation that a visible check decides the outcome, reaches \GuardAddedAcc\%. Paired over the same versions, the four models exceed it by \CthreeAOverGuardSixOne, \CthreeAOverGuardFiveSix, \CthreeAOverGuardClaude, and \CthreeAOverGuardClaudeFourEight{} percentage points. Most of the difference lies in the developer's fixes. Many of them change an existing condition or bound and add no conditional statement, so the rule predicts only \GuardAddedFzero\% of them clean, while the models predict at least \GuardDecFzeroModelMin\% clean. With their renamed copies these fixes contribute \GuardDecFixMin{} to \GuardDecFixMax{} points, the alternative fixes \GuardDecAltMin{} to \GuardDecAltMax, and the decoys that still fault \GuardDecDecoyMin{} to \GuardDecDecoyMax, which the models recognize in \CthreeADecoyFaultClosedMin\% to \CthreeADecoyFaultClosedMax\% of the cases against \GuardAddedDecoyFault\% for the rule. The vulnerable versions change it by \GuardDecVulnMin{} to \GuardDecVulnMax{} points, since the rule always predicts their fault (Table~S9 of the supplementary material). With the concrete input in view, the models thus accept the developer's fix, which the GPT models reject at L1 and L4.

\parag{Benign input.} We run the same driver with a benign input for \BenignFunctions{} functions, for \BenignMechanical{} of them a shortened copy of the trigger input and for \BenignGenerated{} an input in which GPT-6.1-sol changes only the values of the attack, in the input file or in the driver. A benign run is used when execution shows that it reaches the function in $\vv$, for most of them the faulting statement itself, and that $\vv$ and $\ff$ both run clean on it (Section~S22 of the supplementary material). The models predict a clean run of $\vv$ on the benign input for \BenignVzeroCleanMin\% to 100\% of these functions, and for the same $\vv$ they predict the fault on the attack input and a clean run on the benign input in \BenignSwitchedMin\% to \BenignSwitchedMax\% of them, so the predicted outcome follows the input.

\parag{Answer to RQ5.} With the driver and the input in view, the commercial models predict whether a run faults beyond the guard rule, mainly because they predict the developer's fixes clean, and the prediction follows the input.

\section{Memorization and Other Alternative Explanations}
\label{sec:rq4}

The results so far are compatible with partial understanding. This section examines explanations that would produce the same observations without it, first recognition of known vulnerabilities and then robustness to the setup.

\begin{figure}[t]
\centering
\FigFindings
\caption{Margins of the four commercial models over the baseline of each level, in percentage points with 95\% intervals, in the pre-cutoff set and on the CVEs published after their training cutoffs. L1: P-C minus the chance level of the model. L2: share of functions whose cited lines contain the trigger line, minus that of the keyword rule. L3: P-C change from blanking the dynamic slice $D$, minus that from blanking as many random lines. L5: outcome accuracy minus that of the added-guard rule, paired over the same versions. The L1 margin of Claude Opus 5.5 falls from \ChanceGapClaude{} to \PostCutChanceGapClaude{} points, and the L5 margins fall from \CthreeAOverGuardMin{} to \CthreeAOverGuardMax{} points before the cutoffs to \PostCutCthreeAOverGuardMin{} to \PostCutCthreeAOverGuardMax{} after them. Table~S8 of the supplementary material gives all measures.}
\label{fig:findings}
\end{figure}

\parag{Recognition.} A recognition probe shows each commercial model the vulnerable function alone and asks for its project and its CVE. Claude Opus 5.5 names the right CVE for \RecClaudeCVE\% of the pre-cutoff functions, GPT-6.1-sol and GPT-5.6-sol for \RecSixOneCVE\% and \RecFiveSixCVE\%, and Claude Opus 4.8 for \RecClaudeFourEightCVE\%. All four name the right project for most functions, GPT-6.1-sol for \RecSixOneProject\%, and all four separate more pairs when they name the CVE, GPT-6.1-sol \RecSixOneADRcve\% against \RecSixOneADRprojectOnly\% when it names only the project and Claude Opus 5.5 \RecClaudeADRcve\% against \RecClaudeADRprojectOnly\%. For the GPT models recognition covers a small share of the functions, so most of their discrimination comes from functions whose CVE they do not name. The L5 predictions of Claude Opus 5.5 are equally accurate with and without recognition (difference \RecCthreeAClaudeDiff{} points).

\parag{After the training cutoffs.} Figure~\ref{fig:findings} compares each level before and after the cutoffs. None of the four models names the exact CVE of any of the \PostCutStudyCVEs{} post-cutoff functions, while they name the right project for most of them. Localization persists. All four cite the trigger line \PostCutKwGapClosedMin{} to \PostCutKwGapClosedMax{} points more often than the keyword rule, and their cited lines remain closer to the dynamic slice than to the DeepWukong slice. Reliance on the evidence persists as well. Blanking the dynamic slice lowers the P-C of GPT-6.1-sol and now also that of Claude Opus 5.5 against random blanking, by \PostCutMaskDvsRandomSixOne{} and \PostCutMaskDvsRandomClaude{} points, so the verdicts of Claude Opus 5.5 that survived this removal in the pre-cutoff set depend on the dynamic slice on these functions. A single blanked line, however, carries a signal of its own. Blanking only the trigger line raises the P-C of Claude Opus 5.5 by \PostCutMaskVvsNoneClaude{} points, because \PostCutClaudeBlankVNoToYes{} of the \PostCutClaudeBlankVNo{} vulnerable versions it judges not vulnerable when unchanged are judged vulnerable once that line is empty. An empty line where code is expected reads to the model as a sign that something was removed, which is why L3 compares a blanked set with random lines of the same size. The verdicts on edited versions weaken as well. On the execution-confirmed variants of the post-cutoff functions, Claude Opus 5.5 separates \PostCutLFourADRClaude\% of the pairs of real versions, the level of the other three models, so its large L4 advantage before the cutoffs does not carry over (Table~S8 of the supplementary material). Execution prediction weakens. The four models predict \PostCutCthreeAMin\% to \PostCutCthreeAMax\% of the outcomes, about 15 points fewer than in the pre-cutoff set, far above the hidden-body floor, and their paired differences to the added-guard rule, which loses about 5 points, shrink from \CthreeAOverGuardMin{} to \CthreeAOverGuardMax{} points to \PostCutCthreeAOverGuardMin{} to \PostCutCthreeAOverGuardMax{} points, with intervals that all include zero.

\parag{The lead of Claude Opus 5.5.} The pair discrimination of the GPT models stays at its pre-cutoff level, with changes of \DidChangeSixOne{} and \DidChangeFiveSix{} points. The P-C of Claude Opus 5.5 falls from \DidMainClaude\% to \DidPostClaude\% (\DidChangeClaude{} points), and its change minus that of each GPT model is \DidClaudeMinusSixOne{} and \DidClaudeMinusFiveSix{} points. Claude Opus 4.8, the earlier model of the same vendor, also declines, by \DidChangeClaudeFourEight{} points. Within the pre-cutoff set, the P-C of all four models is lower for the \YearRecentFunctions{} functions with a CVE from 2021 or later than for those from 2019 and 2020, without a steady trend over the CVE year (Section~S20 of the supplementary material). Measured against these recent functions, Claude Opus 5.5 still declines by \YearPostMinusRecentClaude{} points, while the changes of the other three models include zero. On the \PostCutFunctions{} post-cutoff pairs with all slices Claude Opus 5.5 remains the only model above chance (\PostCutChanceGapClaude{} points), and its lead over GPT-6.1-sol, \PostCutClaudeVsSixOne{} points on these pairs and \PostCutAllClaudeVsSixOne{} on all \PostCutStudyCVEs{} CVEs, is no longer significant. On all \PostCutStudyCVEs{} CVEs the four models separate a share of the pairs close to that on the \PostCutFunctions{} pairs with all slices, GPT-6.1-sol for example \PostCutADRAllCISixOne\%. Its larger pre-cutoff lead at L1 and L4 and its weaker response to evidence removal are thus consistent with training exposure, and differences between the two sets of functions remain a possible explanation. \parag{Open-weight models after the cutoffs.} The open-weight models stay at or below their chance level on the post-cutoff pairs as well, with an MCC of at most \PostCutOpenMCCmax, so the pre-cutoff result at L1 extends to the later CVEs. Their internal attributions remain at the level of the keyword shortcut, and marking the trigger line in the prompt changes none of their P-C values significantly (Section~S27 of the supplementary material).

\parag{Two versions side by side.} A single version of a function reveals little through its surface, and the surface classifier reaches an AUC of \SurfaceAUC. When both versions are visible, the rule ``the shorter version is the vulnerable one'' is correct for \TcRule\% of the pairs, because fixes mostly add code, so a two-alternative task over the real pair is solvable from length alone. Asked which of the two versions is vulnerable, the commercial models choose correctly for \TcRightClosedMin\% to \TcRightClosedMax\% of the pairs, above this rule, and also for most pairs whose fixed version they label vulnerable when it is shown alone.

\parag{Robustness of the setup.} The vulnerable verdict grows with the length of the function. GPT-6.1-sol judges both versions vulnerable for \LenShortBothVuln\% of the pairs with up to 30 lines and for \LenLongBothVuln\% of those beyond 200 lines, while its P-C falls from \LenShortADR\% to \LenLongADR\%, and Claude Opus 5.5 shows the same trend at a lower level, from \LenShortBothVulnClaude\% to \LenLongBothVulnClaude\%. A longer function offers more statements that could fail for some input, and the screening of Section~\ref{sec:obtaining} found no defect in these fixed versions, so the trend reflects how the models resolve the uncertainty of a long function. Repeated queries give the same answer in all three runs for \RepSameClosedMin\% to \RepSameClosedMax\% of the versions, so differences of a few points between single runs lie within this noise. Removing the trailing newline of PrimeVul changes the answers about as often as asking again, and the P-C differs across families of the Common Weakness Enumeration for no model after a Bonferroni correction (Sections~S5, S6, and~S27 of the supplementary material). The families still differ in how location relates to discrimination. For use-after-free and double-free faults, GPT-6.1-sol cites the trigger line in \CweUafCiteSixOne\% of the functions and separates only \CweUafADRSixOne\% of the pairs, since the faulting release is visible in both versions and the fix acts on the state that decides whether it runs.

\parag{Finding.} Localization and the reliance of the verdicts on the dynamic slice persist on vulnerabilities that the models cannot have seen. The advantage at L5 and the lead of Claude Opus 5.5 at L1 shrink on them, so part of the strongest pre-cutoff results draws on known vulnerabilities or on differences between the two sets of functions.

\section{Discussion}
\label{sec:discussion}

\begin{table}[t]
\centering
\caption{The four findings, the levels behind them, the comparison that decides each, and whether it holds on the CVEs published after the training cutoffs. Numbers are percentage points or percent for the commercial models.}
\label{tab:findings}
\footnotesize
\setlength{\tabcolsep}{3pt}
\begin{tabular}{@{}>{\raggedright\arraybackslash}p{0.22\linewidth}l>{\raggedright\arraybackslash}p{0.37\linewidth}>{\raggedright\arraybackslash}p{0.24\linewidth}@{}}
\toprule
Finding & Levels & Deciding comparison (pre-cutoff set) & After the cutoffs \\
\midrule
F1 Telling a function from its fix is hard & L1 & P-C over chance, at most \ChanceGapClaude{} (Claude Opus 5.5), none for the open-weight models, five of which have more correct than reversed pairs, both versions vulnerable for \BothVulnGptMin\% to \BothVulnGptMax\% (GPT models) & only Claude Opus 5.5 above chance, by \PostCutChanceGapClaude \\
F2 Localization exceeds shortcuts, reliance is partial & L2, L3 & trigger line cited \KwGapClosedMin{} to \KwGapClosedMax{} over the keyword rule, blanking $D$ lowers P-C for three models & margin \PostCutKwGapClosedMin{} to \PostCutKwGapClosedMax, Claude Opus 5.5 now sensitive to $D$ \\
F3 With an input, outcomes are predicted & L4, L5 & accuracy \CthreeAAccMin\% to \CthreeAAccMax\% against \GuardAddedAcc\% for the added-guard rule & accuracy \PostCutCthreeAMin\% to \PostCutCthreeAMax\%, level with the rule \\
F4 Part of the strongest results depends on known CVEs & all & P-C of Claude Opus 5.5 \DidMainClaude\% & \DidPostClaude\%, a change of \YearPostMinusRecentClaude{} against CVEs from 2021 on \\
\bottomrule
\end{tabular}
\end{table}

\parag{What the evidence supports.} Table~\ref{tab:findings} collects the four findings. They separate capabilities that aggregate detection scores conflate. Fault localization succeeds beyond keyword shortcuts and persists after the training cutoffs, while standalone verdicts still flag screened repairs and depend only in part on the evidence a model localizes. With a concrete input, the models reason about whether a check stops a faulting access, yet this advantage shrinks on new vulnerabilities. Open-weight reasoning models sometimes beat the localization baselines while their pair discrimination stays at the chance level, with more correct than reversed pairs, a weak sensitivity to the fix. Whether models rely on the lines they themselves cite remains open, since L3 intervenes on the reference evidence.

\parag{Why localization and discrimination diverge.} The same models that place the fault correctly label its repair vulnerable when they see the repair alone, and they choose the vulnerable version correctly when they see both versions together or predict a clean run of the repair when they see a concrete input. The capability that is missing is thus an absolute judgment of one function without a contrast. Locating the trigger line asks where a fault would occur, while judging a repair asks whether the values that reach that line can still violate its precondition, a question that the surrounding code of one function often leaves open. A detector that answers it with \emph{vulnerable} by default is consistent with the length effect of Section~\ref{sec:rq4} and with the both-vulnerable pairs of Figure~\ref{fig:pairs}.

\parag{What the reference leaves out.} The trigger line and the dynamic slice describe one run that reached the fault. A root cause that acts through allocation state, as the first release in Example~\ref{ex:stale}, or a check that is missing from the function, lies outside the slice, and a citation of such a line counts as an extra line. Overlap with the reference is therefore a lower bound on localization, and the interventions of L3 to L5 complement it with tests that need no line-level reference at all.

\parag{Implications for evaluation.} Three practices follow from the study. Matched pairs with pair-level rates should accompany F1, since F1 rewards indiscriminate \emph{vulnerable} verdicts. Evidence should be compared with a reference derived from the faulting execution and with a keyword rule, since large static references and comparison with random lines alone both inflate apparent alignment. Claims about memorization need data published after the training cutoffs, compared against the most recent data before them.

\parag{Implications for detectors.} The commercial models already cite the trigger line for most functions, and the GPT models still label the developer's fixed version vulnerable for about half of the functions. A concrete input narrows this gap, while supplying slices in the prompt changes it little. Progress may therefore require training signals or tools that expose dependence and execution, such as slice-conditioned objectives or access to a sanitizer run. Concretely, a detector that reviews a patch should see both versions (Section~\ref{sec:rq4}), state the input under which it expects the fault, and have its explanation scored against the faulting run and a keyword rule.

\section{Threats to Validity}
\label{sec:threats}

\parag{Construct validity.} The ladder is one operationalization of understanding. A model may understand a vulnerability and decline to call a single function vulnerable when the outcome depends on callers. The pair-level view and the L5 task, which supplies the driver and the input, address this partially. Stated evidence is a self-report whose faithfulness is not guaranteed~\cite{jacovi2020faithfully,turpin2023unfaithful}, which is why L3 and L4 test the sensitivity of the verdict to the evidence and to confirmed edits.

\parag{Internal validity.} For \NumPatchDerived{} functions the trigger line is patch-derived and rests on the developer's fix, and Table~S6 of the supplementary material reports the two groups separately. Evidence removal modifies only the vulnerable version. The open-weight models fall into two groups by protocol. The five models without a reasoning mode are read at the answer token, so a difference between them and the commercial models reflects both the model and the protocol, while the four reasoning models answer under the protocol of the commercial models. The attribution, prompt paraphrases, and linear head of VulnLLM-R-7B and Qwen3.6-27B come from runs read at the answer token, and gpt-oss-20b and Foundation-Sec-8B-Reasoning have none of the three. Raising the reasoning budget from 8,192 to \ReasonBudgetGptOss{} tokens moved the P-C of gpt-oss-20b from \FirstBudgetADRGptOss\% to \OnADRGptOss\% and that of Qwen3.6-27B from \FirstBudgetADRQwenThreeSix\% to \OnADRQwenThreeSix\%, and neither exceeds its chance level under either budget. Several inputs come from the model families we evaluate. Claude Opus 5.0 wrote the harnesses and inputs, so the drivers that L5 shows to every model and the set of faults in the study both stem from one Claude model, which could favor code that Claude models read well. All drivers share this author, so L5 cannot be split by the source of the driver. Claude Opus 4.8, of the same family, has the lowest L5 accuracy of the four commercial models, which argues against a large advantage of the family. GPT-6.1-sol wrote the alternative fixes and \BenignGenerated{} of the \BenignFunctions{} benign inputs, and Claude Opus 5.5 the post-cutoff decoys, so each of them judges edits of its own, while GPT-5.6-sol and Claude Opus 4.8 wrote none. Commercial models were queried once per prompt, and a repetition estimates how much their answers vary (Section~\ref{sec:rq4}). The strata and the CWE families are exploratory and uncorrected for multiplicity. The keyword list and the surface features are specific choices, and stronger shortcuts may exist. The dynamic slices contain control dependences that sanitizer handler exits introduce in \HcdChanged{} functions, whose effect on the slice comparisons we measured (Section~\ref{sec:obtaining}).

\parag{Conclusion validity.} Functions that share a CVE are resampled together, so the intervals account for their dependence, and paired differences are formed within functions. The strata and the CWE families are exploratory, and the conclusions rest on comparisons that agree across models and across both sets of CVEs. The commercial models were queried once per prompt, and the repetition of Section~\ref{sec:rq4} bounds how much single answers vary.

\parag{External validity.} The functions are C and C++ functions with a reproducible proof of concept and a complete dynamic slice, which favors memory-safety faults and short call chains. The proofs of concept are built with one confirmation method, VulValidate with Claude Opus 5.0, so the set of functions also reflects which faults its dynamic tools can trigger and attribute to the labeled function. Both sets require testing of the fixed version with multiple dynamic analysis tools and varied inputs that reveals no additional defect or vulnerability, and the resulting labels reflect the coverage of that testing. The study covers thirteen LLMs and no fine-tuned detector such as LineVul~\cite{Fu:2022:LineVul} or DeepDFA~\cite{steenhoek2024dataflow}. Applying the same references to those detectors is immediate future work.

\section{Related Work}
\label{sec:related}

\parag{Learning-based vulnerability detection.} Detectors based on code gadgets~\cite{li2018vuldeepecker}, graph neural networks~\cite{zhou2019devign,Reveal}, transformers~\cite{Fu:2022:LineVul}, and data-flow-inspired architectures~\cite{steenhoek2024dataflow} report strong scores on BigVul~\cite{Fan2020}, Devign~\cite{zhou2019devign}, and DiverseVul~\cite{Chen:2023:DiverseVul}. PrimeVul~\cite{ding2024vulnerability} showed that label noise and duplication inflate these scores and introduced paired evaluation. We adopt matched pairs and add an execution-derived reference for each label.

\parag{LLMs as vulnerability detectors.} Recent evaluations find that LLMs struggle to reason about vulnerabilities and are sensitive to prompt and code perturbations~\cite{khare2023understanding,ullah2024llms,steenhoek2024toerr,zhou2024llmvuln}. These studies judge reasoning by comparing explanations with CVE descriptions or human assessment. We compare evidence with the trigger line and the dynamic slice of a faulting run, and we test dependence by removal.

\parag{Robustness and shortcuts.} Detectors change their predictions under semantics-preserving transformations~\cite{risse2024uncovering,rabin2021generalizability} and exploit artifacts of the data~\cite{Broken:2023:Imgrund,arp2022dos,geirhos2020shortcut}. Our tests use shortcuts and controls to measure these effects, and the counterfactual benchmark extends transformation-based testing with variants whose labels are confirmed by execution.

\parag{Explanation evaluation.} Rationale benchmarks compare model evidence with human rationales and test faithfulness by removal~\cite{deyoung2020eraser,jacovi2020faithfully,adebayo2018sanity,jain2019attention}. For vulnerability detectors, explanation methods have been assessed with patch lines or static program structure~\cite{ganz2021explaining,warnecke2020evaluating,hu2023interpreters,cheng2024beyond}. We replace these references with facts of the faulting execution and quantify how static slices differ from them.

\parag{Memorization and evaluation after the cutoff.} Language models can reproduce training data verbatim~\cite{carlini2021extracting}, and code models can recall vulnerable code and its fixes~\cite{Yang2024}, so results on public vulnerability benchmarks may reflect recall. We pair a recognition probe with CVEs published and fixed after the stated training cutoffs, and we compare the post-cutoff results with the most recent CVEs before the cutoffs, which separates a decline over time from a drop at the cutoff.

\parag{Program slicing and execution reasoning.} Static slicing~\cite{weiser1981slicing,tip1995survey} and dynamic slicing~\cite{korel1988dynamic,agrawal1990dynamic,zhang2003precise} are classical tools for fault understanding, and DeepWukong uses static slices as model input~\cite{Cheng:2021:DeepWuKong}. Benchmarks of code execution reasoning test whether LLMs predict program behavior~\cite{liu2024codemind,gu2024cruxeval}. L5 applies this style of evaluation to vulnerabilities with sanitizer-verified outcomes.

\section{Conclusion}
\label{sec:conclusion}

Execution-grounded tests separate what aggregate detection scores conflate. LLM detectors locate the fault of a vulnerable function beyond keyword shortcuts, and this persists on vulnerabilities published after their training. Telling a vulnerable function from its screened repair remains hard, and only partly tied to the evidence the models localize. Given a concrete input, the commercial models predict the outcome of a run beyond a guard rule, an advantage that disappears on the newer vulnerabilities. Matched pairs, execution-derived references, shortcut controls, and post-cutoff data together expose these distinct capabilities.

\bibliography{ECOOP2027_vuln_understanding}

\end{document}